\documentclass[acmsmall,screen=true,review=false]{acmart}
\usepackage{amsmath}
\usepackage{booktabs}%
\usepackage{url}
\usepackage{caption}
\usepackage{colortbl}
\usepackage{CJKutf8}
\usepackage{amsmath}
\usepackage{acronym}
\usepackage{algorithmic}
\usepackage{booktabs}
\usepackage{multirow}

\usepackage{amssymb} 
\usepackage{bbding} 
\usepackage{pifont} 
\usepackage{enumitem}
\usepackage{adjustbox}
\usepackage{xspace}
\usepackage{hyperref}
\usepackage{natbib}
\usepackage{graphicx}
\usepackage{float}
\usepackage{subfigure}
\usepackage{todonotes}
\usepackage{xcolor}
\usepackage{comment}
\usepackage{pgfplots}
\usepackage{pgfplotstable}
\usepackage{makecell}
\usepackage{adjustbox}
\usepackage{pifont}
\usepackage{float}
\usepackage[edges]{forest}
\usepackage{wrapfig}

\AtBeginDocument{%
  \providecommand\BibTeX{{%
    \normalfont B\kern-0.5em{\scshape i\kern-0.25em b}\kern-0.8em\TeX}}}

\definecolor{revise}{HTML}{1834b5}

\setcopyright{acmcopyright}
\copyrightyear{2025}
\acmYear{2025}
\acmDOI{XXXXXXX.XXXXXXX}
\acmJournal{TOIS}

\begin{document}
\title{From Matching to Generation: A Survey on Generative Information Retrieval}

\author{Xiaoxi Li}\orcid{0009-0003-0708-418X}
\email{xiaoxi_li@ruc.edu.cn}
\author{Jiajie Jin}\orcid{0009-0006-4808-1534}
\email{jinjiajie@ruc.edu.cn}
\affiliation{
  \institution{Renmin University of China}
  \city{Beijing}
  \country{China}
}

\author{Yujia Zhou}\orcid{0000-0002-3530-3787}
\email{zhouyujia@mail.tsinghua.edu.cn}
\affiliation{
  \institution{Tsinghua University}
  \city{Beijing}
  \country{China}
}

\author{Yuyao Zhang}\orcid{0009-0009-4632-7934}
\email{2020201710@ruc.edu.cn}
\author{Peitian Zhang}\orcid{0009-0007-1926-7433}
\email{namespace.pt@gmail.com}
\affiliation{
  \institution{Renmin University of China}
  \city{Beijing}
  \country{China}
}

\author{Yutao Zhu}\orcid{0000-0002-9432-3251}
\email{yutaozhu94@gmail.com}
\author{Zhicheng Dou}\orcid{0000-0002-9781-948X}
\authornote{Zhicheng Dou is the corresponding author.}
\email{dou@ruc.edu.cn}
\affiliation{
  \institution{Renmin University of China}
  \city{Beijing}
  \country{China}
}

\renewcommand{\shortauthors}{Li et al.}

\begin{abstract}
Information Retrieval (IR) systems are crucial tools for users to access information, which have long been dominated by traditional methods relying on similarity matching. With the advancement of pre-trained language models, generative information retrieval (GenIR) emerges as a novel paradigm, attracting increasing attention. Based on the form of information provided to users, current research in GenIR can be categorized into two aspects: \textbf{(1) Generative Document Retrieval} (GR) leverages the generative model's parameters for memorizing documents, enabling retrieval by directly generating relevant document identifiers without explicit indexing. \textbf{(2) Reliable Response Generation} employs language models to directly generate information users seek, breaking the limitations of traditional IR in terms of document granularity and relevance matching while offering flexibility, efficiency, and creativity to meet practical needs. This paper aims to systematically review the latest research progress in GenIR. We will summarize the advancements in GR regarding model training and structure, document identifier, incremental learning, etc., as well as progress in reliable response generation in aspects of internal knowledge memorization, external knowledge augmentation, etc. We also review the evaluation, challenges and future developments in GenIR systems. This review aims to offer a comprehensive reference for researchers, encouraging further development in the GenIR field.~\footnote{Github Repository: \url{https://github.com/RUC-NLPIR/GenIR-Survey}}

\end{abstract}




\maketitle

\section{Introduction}
\label{sec:intro}
\begin{figure*}[!t]
    \centering
    \includegraphics[width=0.98\linewidth]{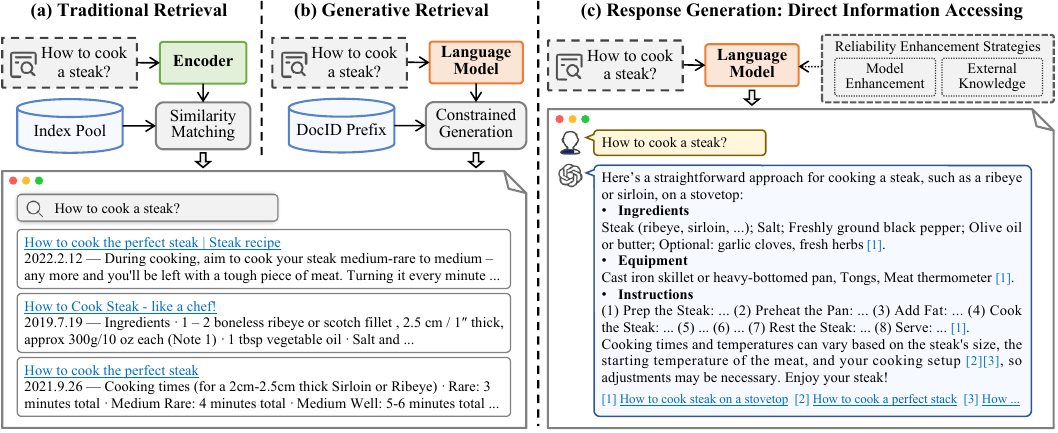}
    \vspace{-2mm}
    \caption{
    Exploring IR Evolution: From Traditional to Generative Methods - This diagram illustrates the shift from traditional similarity-based document matching (a) to GenIR techniques. Current GenIR methods can be categorized into two types: generative retrieval (b), which retrieves documents by directly generating relevant DocIDs constrained by a DocID prefix tree; and response generation (c), which directly generates reliable and user-centric answers.
    }
    \label{fig:compare}
\end{figure*}

Information retrieval (IR) systems are crucial for navigating the vast sea of online information in today's digital landscape. From using search engines such as Google~\cite{Google}, Bing~\cite{Bing}, and Baidu~\cite{chatgpt}, to engaging with question-answering or dialogue systems like ChatGPT~\cite{chatgpt} and Bing Chat~\cite{BingChat}, and discovering content via recommendation platforms like Amazon~\cite{Amazon} and YouTube~\cite{YouTube}, IR technologies are integral to our everyday online experiences. These systems are reliable and play a key role in spreading knowledge and ideas globally. 

Traditional IR systems primarily rely on sparse retrieval methods based on word-level matching. These methods, which include Boolean Retrieval~\cite{boolean_retrieval}, BM25~\cite{bm25}, SPLADE~\cite{splade}, and UniCOIL~\cite{UniCOIL}, establish connections between vocabulary and documents, offering high retrieval efficiency and robust system performance. With the rise of deep learning, dense retrieval methods such as DPR~\cite{dpr} and ANCE~\cite{ance}, based on the bidirectional encoding representations from the BERT model~\cite{bert}, capture the deep semantic information of documents, significantly improving retrieval precision. Although these methods have achieved leaps in accuracy, they rely on large-scale document indices~\cite{faiss, HNSW} and cannot be optimized in an end-to-end way. Moreover, when people search for information, what they really need is a precise and reliable answer. This document ranking list-based IR approach still requires users to spend time summarizing their required answers, which is not ideal enough for information seeking~\cite{metzler2021rethinking}.

With the development of Transformer-based pre-trained language models such as T5~\cite{t5}, BART~\cite{bart}, and GPT~\cite{gpt}, they have demonstrated their strong text generation capabilities. In recent years, large language models (LLMs) have brought about revolutionary changes in the field of AI-generated content (AIGC)~\cite{llm-survey, aigc_survey}. Based on large pre-training corpora and advanced training techniques like RLHF~\cite{rlhf}, LLMs~\cite{chatgpt,llama2,qwen,mistral_7b} have made significant progress in natural language tasks, such as dialogue~\cite{chatgpt,lamda} and question answering~\cite{webglm, webcpm}. The rapid development of LLMs is transforming IR systems, giving rise to a new paradigm of generative information retrieval (GenIR), which achieves IR goals through generative approaches.

As envisioned by Metzler et al.~\cite{metzler2021rethinking}, in order to build an IR system that can respond like a domain expert, the system should not only provide accurate responses but also include source citations to ensure the credibility of the results. To achieve this, GenIR models must possess both sufficient memorized knowledge and the ability to recall the associations between knowledge and source documents, which could be the final goal of GenIR systems. 
Currently, research in GenIR primarily focuses on two main patterns: (1) \textbf{Generative Document Retrieval} (GR), which involves retrieving documents by generating their identifiers; and (2) \textbf{Reliable Response Generation}, which entails directly generating user-centric responses with reliability enhancement strategies. 
Noting that although these two methods have not yet been integrated technically, they represent two primary forms by which IR systems present information to users in generative manners: either by generating lists of document identifiers or by generating reliable and user-centric responses. Figure~\ref{fig:compare} illustrates the difference between these two forms. These strategies are essential to the next generation of information retrieval and constitute the central focus of this survey.

\textbf{Generative document retrieval}, a new retrieval paradigm based on generative models, is garnering increasing attention. This approach leverages the parametric memory of generative models to directly generate document identifiers (DocIDs) related to the documents~\cite{genre, dsi, nci, ultron}. Figure~\ref{fig:compare} illustrates this transition, where traditional IR systems match queries to documents based on an indexed database (Figure~\ref{fig:compare}(a)), while generative methods use language models to retrieve by directly generating relevant document identifiers (Figure~\ref{fig:compare}(b)).
Specifically, GR assigns a unique identifier to each document, which can be numeric-based or text-based, and then trains a generative retrieval model to learn the mapping from queries to the relevant DocIDs. This allows the model to index documents using its internal parameters. During inference, GR models use constrained beam search to limit the generated DocIDs to be valid within the corpus, ranking them based on generation probability to produce a ranked list of DocIDs. This eliminates the need for large-scale document indexes in traditional methods, enabling end-to-end training of the model.

Recent studies on generative retrieval have delved into model training and structure~\cite{dsi, nci, webultron, genrrl, retrollm, ROGER, askari2024generative}, document identifier design~\cite{genre, dsi, genret, asi, valluri2024scaling}, continual learning on dynamic corpora~\cite{dsi++, incdsi, corpusbrain++}, downstream task adaptation~\cite{gere, corpusbrain, corpuslm}, multi-modal generative retrieval~\cite{irgen, gemkr, grace}, and generative recommender systems~\cite{p5, generec, TIGER}. The progress in GR is shifting retrieval systems from matching to generation. 
It has also led to the emergence of workshops~\cite{2023_genir_workshop} and tutorials~\cite{2023_genir_tutorial}. However, there is currently no comprehensive review that systematically organizes the research, challenges, and prospects of this emerging field.

\textbf{Reliable response generation} is also a promising direction in the IR field, offering user-centric and accurate answers that directly meet users' needs. 
Since LLMs are particularly adept at following instructions~\cite{llm-survey}, capable of generating customized responses, and can even cite the knowledge sources~\cite{webgpt, webbrain}, making direct response generation a new and intuitive way to access information~\cite{Sallam2023ChatGPTOR, Gienapp2023EvaluatingGA, White2023FutureSearch, AssistRAG, VIF-RAG}. 
As illustrated in Figure~\ref{fig:compare}, the generative approach marks a significant shift from traditional IR systems, which return a ranked list of documents (as shown in Figure~\ref{fig:compare}(a,b)). Instead, response generation methods (depicted in Figure~\ref{fig:compare}(c)) offer a more dynamic form of information access by directly generating detailed, user-centric responses, thereby providing a richer and more immediate understanding of the information need behind the users' queries.

However, the responses generated by language models may not always be reliable. They have the potential to generate irrelevant answers~\cite{SearchStillMatters}, contradict factual information~\cite{survey_hallu_nlg, survey_hallu_llm}, provide outdated data~\cite{freshllm}, or generate toxic content~\cite{TrustGPT, TrustLLM}. Consequently, these limitations render them unsuitable for many scenarios that require accurate and up-to-date information. 
To address these challenges, the academic community has developed strategies across four key aspects: enhancing internal knowledge~\cite{gpt3, bloom, llama, Lee_dedup_training_data, CaliNet, dola, Sun2020Ernie2.0, ke2302DAP, ROME}; augmenting external knowledge~\cite{rag, self-rag, react, webgpt, FlashRAG, toolformer, search-o1}; generating responses with citation~\cite{According_to_Prompting, IFL, webgpt, LLatrieval, APO}; and improving personal information assistance~\cite{p2bot, SAFARI, xu2023Mental-LLM, RevGAN}. Despite these efforts, there is still a lack of a systematic review that organizes the existing research under this new paradigm of generative information access.

% \definecolor{fill_blue}{RGB}{222, 235, 247}
% \definecolor{fill_red}{RGB}{251, 229, 214}
% \definecolor{fill_green}{RGB}{226, 240, 217}
% \definecolor{fill_yellow}{RGB}{255, 242, 204}

% \definecolor{fill_leaf}{RGB}{240, 240, 240}
% % \definecolor{fill_root}{RGB}{218, 227, 243}
% \definecolor{fill_root}{RGB}{210, 224, 229}
% % \definecolor{fill_root}{RGB}{233, 206, 206}

% \definecolor{fill_blue}{HTML}{e6cecf}
% \definecolor{fill_red}{HTML}{c5dff4}
% \definecolor{fill_green}{RGB}{251, 229, 214}
% \definecolor{fill_yellow}{HTML}{cfe7c4}

% \definecolor{fill_0}{RGB}{205, 232, 251}
% \definecolor{fill_1}{RGB}{249, 237, 231}
% \definecolor{fill_2}{RGB}{238, 243, 237}

% \definecolor{fill_0}{RGB}{218, 232, 244}
% \definecolor{fill_1}{RGB}{252, 235, 224}
% \definecolor{fill_2}{RGB}{231, 243, 230}

\definecolor{fill_0}{RGB}{246, 249, 255}
\definecolor{fill_1}{RGB}{255, 246, 238}
\definecolor{fill_2}{RGB}{244, 249, 241}

\definecolor{draw_0}{RGB}{54, 110, 210}
\definecolor{draw_1}{RGB}{237, 125, 49}
\definecolor{draw_2}{RGB}{112, 173, 71}

\definecolor{fill_leaf}{RGB}{248, 248, 248}
\definecolor{draw-leaf}{RGB}{135, 135, 135}

\tikzstyle{my-box}=[
    rectangle,  
    draw=draw-leaf,
    rounded corners,
    text opacity=1,
    minimum height=1.5em,
    minimum width=5em,
    inner sep=2pt,
    align=center,
    fill opacity=.5,
    line width=0.8pt,
]
\tikzset{
leaf/.style={
my-box,
minimum height=1.5em,
fill=fill_leaf, % Change the RGB values to the desired values
text=black,
align=left,
font=\small,
inner xsep=2pt,
inner ysep=4pt,
line width=1pt
}
}
\begin{figure*}[!t]
\centering
\begin{adjustbox}{width=0.9\textwidth}
\begin{forest}
forked edges,
for tree={
    grow=east,
    reversed=true,
    anchor=base west,
    parent anchor=east,
    child anchor=west,
    base=center,
    font=\large,
    rectangle,
    draw=draw-leaf,
    rounded corners,
    align=left,
    text centered,
    minimum width=5em,
    edge+={darkgray, line width=1pt},
    s sep=3pt,
    inner xsep=2pt,
    inner ysep=3pt,
    line width=1.2pt,
    ver/.style={rotate=90, child anchor=north, parent anchor=south, anchor=center},
},
where level=0{text width=7.5em,fill=fill_0,draw=draw_0,font=\Large,}{},
where level=1{text width=10em,fill=fill_1,draw=draw_1,font=\large,}{},
where level=2{text width=10.4em,fill=fill_2,draw=draw_2,font=\large,}{},
where level=3{font=\normalsize,}{},
% where level=4{text width=7em,font=\normalsize,}{},
[
    GenIR System, minimum height=2.8em%, fill=fill_root
    %%% Chapter: Generative Retrieval
    [ 
        Generative Document\\ Retrieval (Sec \ref{sec:generative_retrieval})%, fill=fill_blue
        [
            Model Training and \\Structure (Sec \ref{sec:model_training_and_structure})%, fill=fill_blue
            [
                \textbf{\textit{Training}} (\ref{sec:gr_training}): DSI~\cite{dsi}{,} DynamicRetriever~\cite{dynamicretriever}{,} NCI~\cite{nci}{,} DSI-QG~\cite{dsiqg}{,}\\ Chen et al.~\cite{chen_understand_dsi}{,} LTRGR~\cite{ltrgr}{,} GenRRL~\cite{genrrl}{,} DGR~\cite{dgr}{,} ListGR~\cite{listgr}{,}, leaf, text width=31.8em
            ]
            [
                \textbf{\textit{Structure}} (\ref{sec:gr_structure}): NCI~\cite{nci}{,} TOME~\cite{tome}{,} NP Decoding~\cite{np-decoding}{,} MEVI~\cite{mevi}{,}\\ DiffusionRet~\cite{diffusion-ret}{,} GDR~\cite{gdr}{,} Self-Retrieval~\cite{self-retrieval}{,} PAG~\cite{PAG}, leaf, text width=31.8em
            ]
        ]
        [
            Document Identifier\\ (Sec \ref{sec:docid_design})%, fill=fill_blue
            [
                \textbf{\textit{Numeric}} (\ref{sec:numeric_docid}): DSI~\cite{dsi}{,} DynamicRetriever~\cite{dynamicretriever}{,} Ultron~\cite{ultron}{,} GenRet~\cite{genret}{,}\\ Tied-Atomic~\cite{gen_as_dense}{,} MEVI~\cite{mevi}{,} LMIndexer~\cite{lmindexer}{,} ASI~\cite{asi}{,} RIPOR~\cite{ripor}, leaf, text width=31.8em
            ]
            [
                \textbf{\textit{Text}} (\ref{sec:text_docid}): GENRE~\cite{genre}{,} SEAL~\cite{seal}{,} Ultron~\cite{ultron}{,} LLM-URL~\cite{llm-url}{,}\\ UGR~\cite{ugr}{,} MINDER~\cite{minder}{,} AutoTSG~\cite{autotsg}{,} SE-DSI~\cite{autotsg}{,} NOVO~\cite{novo}{,} GLEN~\cite{glen}, leaf, text width=31.8em
            ]
        ]
        [
            Incremental Learning\\ (Sec \ref{sec:continual_learning})%, fill=fill_blue
            [
                DSI++~\cite{dsi++}{,} IncDSI~\cite{incdsi}{,} CLEVER~\cite{clever}{,} CorpusBrain++~\cite{corpusbrain++}, leaf, text width=31.8em
            ]
        ]
        [
            Downstream Task \\Adaptation (Sec \ref{sec:downstream_adaption})%, fill=fill_blue
            [
                \textbf{\textit{Separate Training}} (\ref{sec:gr_seperate_train}): GERE~\cite{gere}{,} CorpusBrain~\cite{corpusbrain}{,} GMR~\cite{gmr}{,} DearDR~\cite{deardr}{,}\\ CodeDSI~\cite{codedsi}{,} UGR~\cite{ugr}{,} GCoQA~\cite{GCoQA}{,} Re3val~\cite{re3val}, leaf, text width=31.8em
            ]
            [
                \textbf{\textit{Joint Training}} (\ref{sec:gr_joint_train}): UniGen~\cite{unigen}{,} CorpusLM~\cite{corpuslm}{,} RetroLLM~\cite{retrollm}, leaf, text width=31.8em
            ]
            [
                \textbf{\textit{Multi-Modal}} (\ref{sec:multi_modal}): IRGen~\cite{irgen}{,} GeMKR~\cite{gemkr}{,} GRACE~\cite{grace}, leaf, text width=31.8em
            ]
            [
                \textbf{\textit{Generative Recommender Systems}} (\ref{sec:recommender}):
                P5~\cite{p5}{,} TIGER~\cite{TIGER}{,} SEATER~\cite{seater}{,}\\ IDGenRec~\cite{IDGenRec}{,} LC-Rec~\cite{LCRec}{,} ColaRec~\cite{ColaRec}, leaf, text width=31.8em
                % GeneRec~\cite{generec} GPT4Rec~\cite{GPT4Rec}{,}
            ]
        ]
        % [
        %     Generative Recommender\\ Systems (Sec \ref{sec:recommender}), text width=11.9em%, fill=fill_blue
        %     [
        %         P5~\cite{p5}{,} TIGER~\cite{TIGER}{,} SEATER~\cite{seater}{,} IDGenRec~\cite{IDGenRec}{,} LC-Rec~\cite{LCRec}{,}\\ ColaRec~\cite{ColaRec}, leaf, text width=30.3em
        %         % GeneRec~\cite{generec} GPT4Rec~\cite{GPT4Rec}{,}
        %     ]
        % ]
    ]
    %%% Chapter: Response Generation
    [ 
        Reliable Response\\ Generation (Sec \ref{sec:response_generation}) %, fill=fill_red
        [
            Internal Knowledge\\ Memorization (Sec \ref{sec:model_internal_enhance}) %, text width=10em, fill=fill_red
            [
                \textbf{\textit{Structure}} (\ref{sec:structural_enhance}): 
                \textit{Model Scaling:} GPT3~\cite{gpt3}{,} BLOOM~\cite{bloom}{,} LLaMA~\cite{llama}{,}\\ \textit{Model Structure:} PaLM~\cite{palm}{,} Mixtral 8x7B~\cite{mixtral8x7b}, leaf, text width=31.8em
            ]
            [
                \textbf{\textit{Training and Inference}} (\ref{sec:training_inference_enhance}): 
                \textit{Training:} Sadeq et al.~\cite{Sadeq_improve_fact} FactTune~\cite{FactTune}{;}\\ 
                \textit{Inference:} GenRead~\cite{genread}{,} RECITE~\cite{recite}{,} DoLa~\cite{dola}, leaf, text width=31.8em%, fill=fill_red
            ]
            [
                \textbf{\textit{Knowledge Updating}} (\ref{sec:knowledge_updating}): 
                % \textit{Incremental Pre-training:} Ernie2.0~\cite{Sun2020Ernie2.0}{,} DAS~\cite{ke2302DAS};\\
                % \textit{Incremental Fine-tuning:} Progressive prompts~\cite{Razdaibiedina2301Progressive_prompts}{,} DynaInst~\cite{mok2023DynaInst}{,} O-LoRA~\cite{wang2310O-LoRA}
                \textit{Incremental Learning:} Ernie 2.0~\cite{Sun2020Ernie2.0}{,} DAP~\cite{ke2302DAP};\\
                DynaInst~\cite{mok2023DynaInst}{,}
                \textit{Knowledge Editing:} KE~\cite{KE}{,} MEND~\cite{MEND}{,} ROME~\cite{ROME}
                , leaf, text width=31.8em
            ]
            % [
            %     \textbf{\textit{Training}} (\ref{sec:training_enhance}): 
            %     \textit{Training Data:} Lee et al.~\cite{Lee_dedup_training_data}{,} Sadeq et al.~\cite{Sadeq_improve_fact}{;}\\ 
            %     \textit{Training Methods:} CaliNet~\cite{CaliNet}{,} LIMA~\cite{LIMA}{,} FactTune~\cite{FactTune}, leaf, text width=31.8em%, fill=fill_red
            % ]
            % [
            %     \textbf{\textit{Inference}} (\ref{sec:inference_enhance}): \textit{Prompt Engineering:} GenRead~\cite{genread}{,} RECITE~\cite{recite}{;}\\ \textit{Decoding Strategies:} Nucleus Sampling~\cite{nucleus_sampling}{,} DoLa~\cite{dola}, leaf, text width=31.8em
            % ]
        ]
        [
            External Knowledge\\ Augmentation (Sec \ref{sec:external_knowledge_enhance}) %, text width=10em, fill=fill_red
            [
                \textbf{\textit{Retrieval}} (\ref{sec:retrieval_augmentation}): 
                \textit{Sequential:} RAG~\cite{rag}{,} RRR~\cite{rrr}{,} ARL2~\cite{arl2};\\
                \textit{Branching:} TOC~\cite{toc}{,} BlendFilter~\cite{blendfilter}{,} REPLUG~\cite{replug};\\
                \textit{Conditional:} SKR~\cite{skr}{,} Self-DC~\cite{selfdc}{,} Rowen~\cite{rowen};\\
                \textit{Loop:} Iter-RetGen~\cite{iterretgen}{,} IR-COT~\cite{ircot}{,} FLARE~\cite{flare}{,} Self-RAG~\cite {self-rag}{,} Search-o1~\cite{search-o1}
                , leaf, text width=31.8em
            ]
            [
                \textbf{\textit{Tool}} (\ref{sec:tool_augmentation}): 
                \textit{Search Engine:} ReAct~\cite{react}{,} WebGPT~\cite{webgpt};\\
                \textit{Knowledge Graph:} StructGPT~\cite{structgpt}{,} ToG~\cite{tog}{,} RoG~\cite{rog};\\
                \textit{API-based Tools:} Toolformer~\cite{toolformer}{,} ToolLLM~\cite{ToolLLM}{,} AssistGPT~\cite{assistgpt};\\
                \textit{Model-based Tools:} HuggingGPT~\cite{hugginggpt}{,} Visual ChatGPT \cite{visualgpt}
                , leaf, text width=31.8em
            ]
        ]
        [
            Generating Response\\ with Citation (Sec \ref{sec:response_with_citation}) %, text width=10em, fill=fill_red
            [
                \textbf{\textit{Direct Citation}} (\ref{sec:direct_citation}): 
                According-to Prompting~\cite{According_to_Prompting}{,} IFL~\cite{IFL}{,} Fierro et al.~\cite{Fierro2024Plan_and_Citation}{,}\\ Credible without Credit~\cite{Credible_without_Credit}{,} 1-PAGER~\cite{1-PAGER}{,} Khalifa et al.~\cite{Khalifa2024SourceAwareTE}
                , leaf, text width=31.8em
            ]
            [
                \textbf{\textit{Retrieval-based Citation}} (\ref{sec:retrieval_based_citation}): 
                WebGPT~\cite{webgpt}{,} WebBrain~\cite{webbrain}{,} RARR~\cite{RARR}{,}\\ 
                SearChain~\cite{Search-in-the-Chain}{,} LLatrieval~\cite{alce}{,} VTG~\cite{VTG}{,} CEG~\cite{CEG}{,} APO~\cite{APO}
                , leaf, text width=31.8em
            ]
        ]
        [
            Personal Information \\Assistant (Sec \ref{sec:personal_information_assistant}) %, text width=10em, fill=fill_red
            [
                \textbf{\textit{Personalized Dialogue}} (\ref{sec:direct_personalized}):
                Zhang et al.~\cite{Zhang2018I_have_a_dog}{,} $\mathcal{P}^{2}$Bot~\cite{p2bot}{,} Wu et al.~\cite{Wu2021Personalized_Response_Generation}{,} \\ SAFARI~\cite{SAFARI}{,} Personalized Soups~\cite{PersonalizedSoups}{,} OPPU~\cite{OPPU}
                , leaf, text width=31.8em
            ]
            [
                \textbf{\textit{Domain-specific}} (\ref{sec:rag_personalized}):
                \textit{Healthcare:} Zhongjing~\cite{liu2023Pharmacygpt}{,} Mental-LLM~\cite{xu2023Mental-LLM}{;}\\
                \textit{Academic:} RevGAN~\cite{RevGAN}{,} Pearl~\cite{Mysore2311Pearl}{;}
                \textit{Education:} EduChat~\cite{dan2023Educhat}{;} 
                \textit{Recipe}{,} \textit{Robot}{,} etc.
                , leaf, text width=31.8em
            ]
        ]
    ]
    %%% Chapter: Evaluation
    [
        Evaluation (Sec \ref{sec:evaluation})%, fill=fill_green
        [
            Generative Document\\ Retrieval (Sec \ref{sec:eval_gr})%, text width=9.7em%, fill=fill_green
            [
                \textbf{\textit{Metrics}} (\ref{sec:gr_metrics}): Recall{,} MRR~\cite{mrr}{,} R-Precision{,} MAP{,} nDCG~\cite{ndcg}, leaf, text width=31.8em
            ]
            [
                \textbf{\textit{Benchmarks}} (\ref{sec:gr_benchmarks}): MS MARCO~\cite{ms_marco}{,} NQ~\cite{nq}{,} TriviaQA~\cite{triviaqa}{,} KILT~\cite{kilt}{,}\\ TREC DL 19 \& 20~\cite{trec_19_dl, trec_20_dl}{,} DynamicIR~\cite{dynamicir}{,} Liu et al.~\cite{gr-ood}{,} ExcluIR~\cite{ExcluIR}, leaf, text width=31.8em
            ]
            [
                \textbf{\textit{Analysis}} (\ref{sec:gr_analysis}): Chen et al.~\cite{chen_understand_dsi}{,} Pradeep et al.~\cite{pradeep_how_does}{,} Liu et al.~\cite{gr-ood}{,}\\ Wu et al.~\cite{wu2024gr_as_mvdr}, leaf, text width=31.8em
            ]
            [
                \textbf{\textit{Experiments}} (\ref{sec:gr_exp}): Performance Comparison on MS MARCO~\cite{ms_marco}{,} NQ~\cite{nq}\\ and KILT~\cite{kilt} Benchmarks, leaf, text width=31.8em
            ]
        ]
        [
            Reliable Response\\ Generation (Sec \ref{sec:eval_response})%, text width=9.7em%, fill=fill_green
            [
                \textbf{\textit{Metrics}} (\ref{sec:response_metrics}): \textit{Rule-based:} EM{,} BLEU~\cite{bleu}{,} ROUGE~\cite{rouge}{,} Perplexity;\\ 
                \textit{Model-based:} BERTScore~\cite{bertscore}{,} BLEURT~\cite{BLEURT}{,} GPTScore~\cite{gptscore}{,} FActScore~\cite{FActScore};\\ 
                \textit{Human Evaluation:} Comprehensibility{,} Relevance{,} Fluency
                , leaf, text width=31.8em
            ]
            [
                \textbf{\textit{Benchmarks}} (\ref{sec:response_benchmarks}):
                \textit{General:} MMLU~\cite{MMLU}{,} BIG-bench~\cite{BIG-bench}{,} LLM-Eval~\cite{LLM-Eval};\\ 
                \textit{Tool:} API-Bank~\cite{API-Bank}{,} ToolBench~\cite{ToolLLM};\\
                \textit{Factuality:} TruthfulQA~\cite{TruthfulQA}{,} ALCE~\cite{alce}{,} HaluEval~\cite{halueval};\\
                \textit{Real Time:} RealTime QA~\cite{RealTimeQA}{,} FreshQA~\cite{freshllm};\\
                \textit{Trustworthy:} SafetyBench~\cite{SafetyBench}{,} TrustGPT~\cite{TrustGPT}{,} TrustLLM~\cite{TrustLLM}
                , leaf, text width=31.8em
            ]
            % [
            %     \textbf{\textit{Analysis}} (\ref{sec:response_analysis}): Kadavath et al.~\cite{Kadavath_lm_know}{,} SelfCheckGPT~\cite{SelfCheckGPT}{,} FActScore~\cite{FActScore}{,}\\ RAGAS~\cite{RAGAS}{,} TrustLLM~\cite{TrustLLM}, leaf, text width=31.8em
            % ]
        ]
    ]
    %%% Chapter: Challenges
    [
        Challenges and \\Prospects (Sec \ref{sec:challenge})%, fill=fill_yellow
        [
            Generative Document\\ Retrieval (Sec \ref{sec:challenge_gr})%, text width=9.7em%, fill=fill_yellow
            [
                \textbf{\textit{Scalability}} (\ref{subsec:scalability_issues}){;} 
                \textbf{\textit{Dynamic Corpora}} (\ref{subsec:handling_dynamic_corpora}){;} 
                \textbf{\textit{Document Representation}} (\ref{subsec:document_representation}){;}\\ 
                \textbf{\textit{Efficiency}} (\ref{subsec:efficiency_concerns}){;} \textbf{\textit{Multi-modal}} (\ref{subsec:challenge_multi_modal}), leaf, text width=31.8em
            ]
        ]
        [
            Reliable Response\\ Generation (Sec \ref{sec:challenge_response})%, text width=9.7em%, fill=fill_yellow
            [
                \textbf{\textit{Accuracy and Factuality}} (\ref{subsec:improving_accuracy}){;} 
                \textbf{\textit{Real-time Property}} (\ref{subsec:real_time}){;}\\ 
                \textbf{\textit{Bias and Fairness}} (\ref{subsec:bias}){;} 
                \textbf{\textit{Privacy and Security}} (\ref{subsec:data_privacy})
                , leaf, text width=31.8em
            ]
        ]
        [
            Unified Framework\\ (Sec \ref{sec:unified_framework})%, text width=9.7em%, fill=fill_yellow
            [
                \textbf{\textit{Unified Framework for Retrieval and Generation}} (\ref{subsec:unified_retrieval_generation}){;} \\
                \textbf{\textit{Towards End2end Framework for Various IR Tasks}} (\ref{subsec:end2end_ir_framework})
                , leaf, text width=31.8em
            ]
        ]
    ]
]
\end{forest}
\end{adjustbox}
\vspace{-2mm}
\caption{Taxonomy of research on generative information retrieval: investigating generative document retrieval, reliable response generation, evaluation, challenges and prospects.}
\label{fig:overview_tree}
\end{figure*}
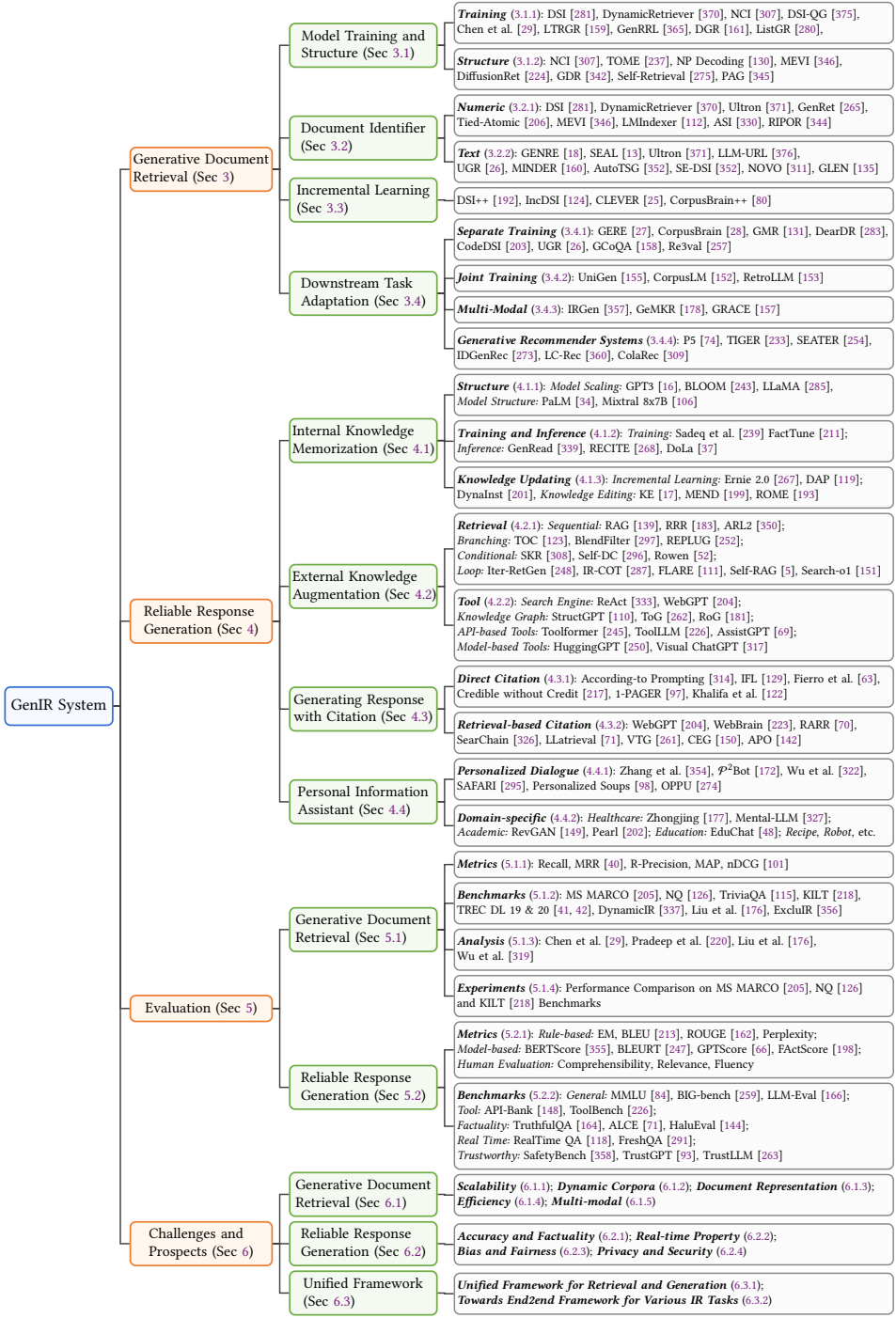

This review will systematically review the latest research progress and future developments in the field of GenIR, as shown in Figure~\ref{fig:overview_tree}, which displays the classification of research related to the GenIR system. We will introduce background knowledge in Section~\ref{sec:background}, generative document retrieval technologies in Section~\ref{sec:generative_retrieval}, direct information accessing with generative language models in Section~\ref{sec:response_generation}, evaluation in Section~\ref{sec:evaluation}, current challenges and future directions in Section~\ref{sec:challenge}, respectively. Section~\ref{sec:conclusion} will summarize the content of this review. This article is the first to systematically organize the research, evaluation, challenges and prospects of generative IR, while also looking forward to the potential and importance of GenIR's future development. Through this review, readers will gain a deep understanding of the latest progress in developing GenIR systems and how it shapes the future of information access. 
The main contribution of this survey is summarized as follows:
\begin{itemize}[leftmargin=*]
\item \textbf{First comprehensive survey on generative information retrieval (GenIR):} This survey is the first to comprehensively organize the techniques, evaluation, challenges, and prospects on the emerging field of GenIR, providing a deep understanding of the latest progress in developing GenIR systems and its future in shaping information access.
\item \textbf{Systematic categorization and in-depth analysis:} The survey offers a systematic categorization of research related to GenIR systems, including generative document retrieval, reliable response generation. It provides an in-depth analysis of each category, covering model training and structure, document identifier, etc. in generative document retrieval; internal knowledge memorization, external knowledge enhancement, etc. for reliable response generation.
\item \textbf{Comprehensive review of evaluation metrics and benchmarks:} The survey reviews a range of widely used evaluation metrics and benchmark datasets for accessing GenIR methods, alongside analysis on the effectiveness and weaknesses of existing GenIR methods.
\item \textbf{Discussions of current challenges and future directions:} The survey identifies and discusses the current challenges faced in the GenIR field. We also provide potential solutions for each challenge and outline future research directions for building GenIR systems. 
\end{itemize}

\section{Background and Preliminaries}
\label{sec:background}

Information retrieval techniques aim at efficiently obtaining, processing, and understanding information from massive data. Technological advancements have continuously driven the evolution of these methods: from early keyword-based sparse retrieval to deep learning-based dense retrieval, and more recently, to generative retrieval, large language models, and their augmentation techniques. Each advancement enhances retrieval accuracy and efficiency, catering to the complex and diverse query needs of users.

\subsection{Traditional Information Retrieval}

\textbf{Sparse Retrieval.} 
In the field of traditional information retrieval, sparse retrieval techniques implement fast and accurate document retrieval through the inverted index method. Inverted indexing technology maps each unique term to a list of all documents containing that term, providing an efficient means for information retrieval in large document collections. Among these methods, TF-IDF (Term Frequency-Inverse Document Frequency)~\cite{tfidf} is a particularly important statistical tool used to assess the importance of a word in a document collection, thereby widely applied in various traditional retrieval systems.

The core of sparse retrieval technology lies in evaluating the relevance between documents and user queries. Specifically, given a document collection \(\mathcal{D}\) and a user query \(q\), traditional information retrieval systems identify and retrieve information by calculating the relevance \(\mathcal{R}\) between document \(d\) and query \(q\). This relevance evaluation typically relies on the similarity measure between document \(d\) and query \(q\), as shown below:
\begin{equation}
\mathcal{R}(q,d) = \sum\nolimits_{t \in q \cap d} \text{tf-idf}(t, d) \cdot \text{tf-idf}(t, q),
\end{equation}
where \(t\) represents the terms common to both query \(q\) and document \(d\), and \(\text{tf-idf}(t, d)\) and \(\text{tf-idf}(t, q)\) represent the TF-IDF weights of term \(t\) in document \(d\) and query \(q\), respectively. Although sparse retrieval methods like TF-IDF~\cite{tfidf} and BM25~\cite{bm25} excel at fast retrieval, it struggles with complex queries involving synonyms, specialized terms, or context, as term matching and TF-IDF may not fully meet users' information needs~\cite{compare_sparse_dense}.

\textbf{Dense Retrieval.} 
\label{sec:background_gr}
The advent of pre-trained language models like BERT~\cite{bert} has revolutionized information retrieval, leading to the development of dense retrieval methods, like DPR~\cite{dpr}, ANCE~\cite{ance}, E5~\cite{e5}, SimLM~\cite{simlm}. Unlike traditional sparse retrieval, these methods leverage Transformer-based encoders to create dense vector representations for both queries and documents. This approach enhances the capability to grasp the underlying semantics, thereby improving retrieval accuracy.

The core of dense retrieval lies in converting documents and queries into vector representations. Given document \(d\) and query \(q\) and their vector representations \(\mathbf{v}_q\), each document \(d\) is transformed into a dense vector \(\mathbf{v}_d\) through a pre-trained language model, similarly, query \(q\) is transformed into vector \(\mathbf{v}_q\). Specifically, we can use encoder functions \(E_d(\cdot)\) and \(E_q(\cdot)\) to represent the encoding process for documents and queries, respectively:
\begin{equation}
\mathbf{v}_d = E_d(d), \quad \mathbf{v}_q = E_q(q),
\label{eq:background_gr}
\end{equation}
where \(E_d(\cdot)\) and \(E_q(\cdot)\) can be the same model or different models optimized for specific tasks.

Dense retrieval methods evaluate relevance by calculating the similarity between the query vector and document vector, which can be measured by cosine similarity, expressed as follows:
\begin{equation}
\mathcal{R}(q,d) = \text{cos}(\mathbf{v}_q, \mathbf{v}_d) = \frac{\mathbf{v}_q \cdot \mathbf{v}_d}{|\mathbf{v}_q| |\mathbf{v}_d|},
\end{equation}
where \(\mathbf{v}_q \cdot \mathbf{v}_d\) represents the dot product of query vector \(\mathbf{v}_q\) and document vector \(\mathbf{v}_d\), and \(|\mathbf{v}_q|\) and \(|\mathbf{v}_d|\) respectively represent the magnitudes of the query and document vector. Finally, documents are ranked based on these similarity scores to identify the most relevant ones for the user. 

% \textbf{Model Optimization.} Dense retrieval models are typically optimized using contrastive learning, which encourages the model to bring the embeddings of relevant query-document pairs closer while pushing apart those of irrelevant pairs. A common loss function employed is the InfoNCE loss, defined as:
% \begin{equation}
% \mathcal{L} = -\log \frac{\exp(\text{sim}(\mathbf{v}_q, \mathbf{v}_d^+))}{\exp(\text{sim}(\mathbf{v}_q, \mathbf{v}_d^+)) + \sum\nolimits_{d^-} \exp(\text{sim}(\mathbf{v}_q, \mathbf{v}_{d^-}))},
% \end{equation}
% where \(\mathbf{v}_q\) and \(\mathbf{v}_d^+\) are the vectors for the query and a positive document, respectively, and \(\mathbf{v}_{d^-}\) represents vectors for negative documents. The similarity function \(\text{sim}(\cdot, \cdot)\) is typically cosine similarity. This loss function effectively trains the model to distinguish relevant documents from a pool of negatives, enhancing retrieval accuracy.

\subsection{Generative Retrieval}

With the significant progress of language models, {generative retrieval} has emerged as a new direction in the field of information retrieval~\cite{metzler2021rethinking, dsi, xu2025survey}. Unlike traditional index-based retrieval methods, generative retrieval relies on pre-trained generative language models, such as T5~\cite{t5} and BART~\cite{bart}, to directly generate document identifiers (DocIDs) relevant to the query, thereby achieving end-to-end retrieval without relying on large-scale pre-built document indices.

\textbf{DocID Construction and Prefix Constraints.}
To facilitate generative retrieval, each document \( d \) in the corpus \( \mathcal{D} = \{d_1, d_2, \dots, d_N\} \) is assigned a unique document identifier \( d' \), forming the set \( \mathcal{D'} = \{d'_1, d'_2, \dots, d'_N\} \). This mapping is typically established via a bijective %\footnote{Noting that this mapping does not absolutely require a unique identifier for each document; that is, multiple documents can share the same DocID. However, the duplication rate of DocIDs should usually be less than 5\% to ensure accuracy.} 
function \( \phi: \mathcal{D} \to \mathcal{D'} \), ensuring that:
\begin{equation}
\phi(d_i) = d'_i, \quad \forall d_i \in \mathcal{D}.
\end{equation}
To enable the language model to generate only valid DocIDs during inference, we construct \textit{prefix constraints} based on \( \mathcal{D'} \). This is typically implemented using a trie (prefix tree), where each path from the root to a leaf node corresponds to a valid DocID.

\textbf{Constrained Beam Search.}
Given a query \( q \), the generative retrieval model aims to generate the top-\( k \) DocIDs that are most relevant to \( q \). The language model \( P(\cdot|q; \theta) \) generates DocIDs token by token, guided by the prefix constraints. At each decoding step \( i \), only those tokens that extend the current partial sequence \( d'_{<i} \) into a valid prefix of some DocIDs in \( \mathcal{D'} \) are considered. Formally, the set of allowable next tokens is:
\begin{equation}
\mathcal{V}(d'_{<i}) = \{ v \mid \exists d' \in \mathcal{D'} \text{ such that } d'_{<i}v \text{ is a prefix of } d' \}.
\end{equation}
By employing \textit{constrained beam search}, the model efficiently explores the space of valid DocIDs, maintaining a beam of the most probable sequences at each decoding step while adhering to the DocID prefix constraints.

\textbf{Document Relevance.}
The relevance between the query \( q \) and a document \( d \) is quantified by the probability of generating its corresponding DocID \( d' \) given \( q \). This is computed as:
\begin{equation}
\mathcal{R}(q, d) = P(d' | q; \theta) = \prod\nolimits_{i=1}^{T} P(d'_i \mid d'_{<i}, q; \theta),
\end{equation}
where \( T \) is the length of the DocID \( d' \) in tokens, \( d'_i \) is the token at position \( i \), and \( d'_{<i} \) denotes the sequence of tokens generated before position \( i \). The constrained beam search produces a ranked list of top-\( k \) DocIDs \( \{d'^{(1)}, d'^{(2)}, \dots, d'^{(k)}\} \) based on their generation probabilities \( \{\mathcal{R}(q, d^{(1)}), \mathcal{R}(q, d^{(2)}),\) \(\dots, \mathcal{R}(q, d^{(k)})\} \). The corresponding documents \( \{d^{(1)}, d^{(2)}, \dots, d^{(k)}\} \) are then considered the most relevant to the query \( q \).

\textbf{Model Optimization.} Generative retrieval models are typically optimized using cross-entropy loss, which measures the discrepancy between the generated DocID sequence and the ground truth DocID. Given a query \( q \) and its corresponding DocID \( d' \), the cross-entropy loss is defined as:
\begin{equation}
\mathcal{L} = - \sum\nolimits_{i=1}^{T} \log P(d'_i \mid d'_{<i}, q; \theta),
\end{equation}
where \( T \) is the length of the DocID in tokens, \( d'_i \) is the token at position \( i \), and \( d'_{<i} \) denotes the sequence of tokens generated before position \( i \). This loss function encourages the model to learn the association between query and labeled DocID sequence.

This approach allows the generative retrieval model to produce a relevance-ordered list of documents without relying on traditional indexing structures. The core of this approach lies in leveraging the language model's capability to generate DocID sequences within prefix constraints. This section discusses the simplest generative retrieval method. In Section~\ref{sec:generative_retrieval}, we will delve into advanced methods from multiple perspectives, including model architectures, training strategies, and DocID design, to further enhance retrieval performance across various scenarios.

\subsection{Large Language Models}

The evolution of Large Language Models (LLMs) marks a significant leap in natural language processing (NLP), rooted from early statistical and neural network-based language models~\cite{llm4ir}. These models, through pre-training on vast text corpora, learned deep semantic features of language, greatly enriching the understanding of text. Subsequently, generative language models, most notably the GPT series~\cite{gpt, gpt2, gpt3}, significantly improved text generation and understanding capabilities with improved model size and number of parameters.

LLMs can be mainly divided into two categories: encoder-decoder models and decoder-only models. Encoder-decoder models, like T5~\cite{t5} and BART~\cite{bart}, convert input text into vector representations through their encoder, then the decoder generates output text based on these representations. This model perspective treats various NLP tasks as text-to-text conversion problems, solving them through text generation. On the other hand, decoder-only models, like the GPT~\cite{gpt} and GPT-2~\cite{gpt2}, rely entirely on the Transformer decoder, generating text step by step through the self-attention mechanism. The introduction of GPT-3~\cite{gpt3}, with its 175 billion parameters, marked a significant milestone in this field and led to the creation of models like InstructGPT~\cite{InstructGPT}, Falcon~\cite{falcon}, PaLM~\cite{palm} and Llama series~\cite{llama, llama2, llama3}. These models, all using a decoder-only architecture, trained on large-scale datasets, have shown astonishing language processing capabilities~\cite{llm-survey}.

For information retrieval tasks, large language models (LLMs) play a crucial role in directly generating the exact information users seek~\cite{llm4ir, FullRank, DPA-RAG}. This capability marks a significant step towards a new era of generative information retrieval. In this era, the retrieval process is not solely about locating existing information but also about creating new content that meets the specific needs of users. This feature is especially advantageous in situations where users might not know how to phrase their queries or when they are in search of complex and highly personalized information, scenarios where traditional matching-based methods fall short.

\subsection{Augmented Language Models}

Despite the advances of LLMs, they still face significant challenges such as hallucination, particularly in complex tasks or those requiring access to long-tail or real-time information~\cite{llm-survey, survey_hallu_llm}. To address these issues, retrieval augmentation and tool augmentation have emerged as effective strategies. Retrieval augmentation involves integrating external knowledge sources into the language model's workflow. This integration allows the model to access up-to-date and accurate information during the generation process, thereby grounding its responses in verified data and reducing the likelihood of hallucinations~\cite{rag, replug, HtmlRAG}. Tool augmentation, on the other hand, extends the capabilities of LLMs by incorporating specialized tools or APIs that can perform specific functions like mathematical computations, data retrieval, or executing predefined commands~\cite{toolformer, toolalpaca, ToolLLM}. With retrieval and tool augmentations, language models can provide more precise and contextually relevant responses, thereby improving factuality and functionality in practical applications.

Moreover, due to the aforementioned issue of hallucinations, the responses generated by LLMs are often considered unreliable because users are unaware of the sources behind the generated content, making it difficult to verify its accuracy. To enhance the credibility of responses, some studies have focused on generating responses with citations~\cite{survey_LLM_attribution, Attribute_First_then_Generate, webgpt}. This approach involves enabling language models to cite the source documents of their generated content, thereby increasing the trustworthiness of the responses. All these methods are effective strategies for improving both the quality and reliability of language model outputs and are essential technologies for building the next generation of generative information retrieval systems.

\section{Generative Document Retrieval: From Similarity Matching to Generating Document Identifiers}
\label{sec:generative_retrieval}

\begin{figure*}[!t]
    \centering
    \includegraphics[width=0.998\linewidth]{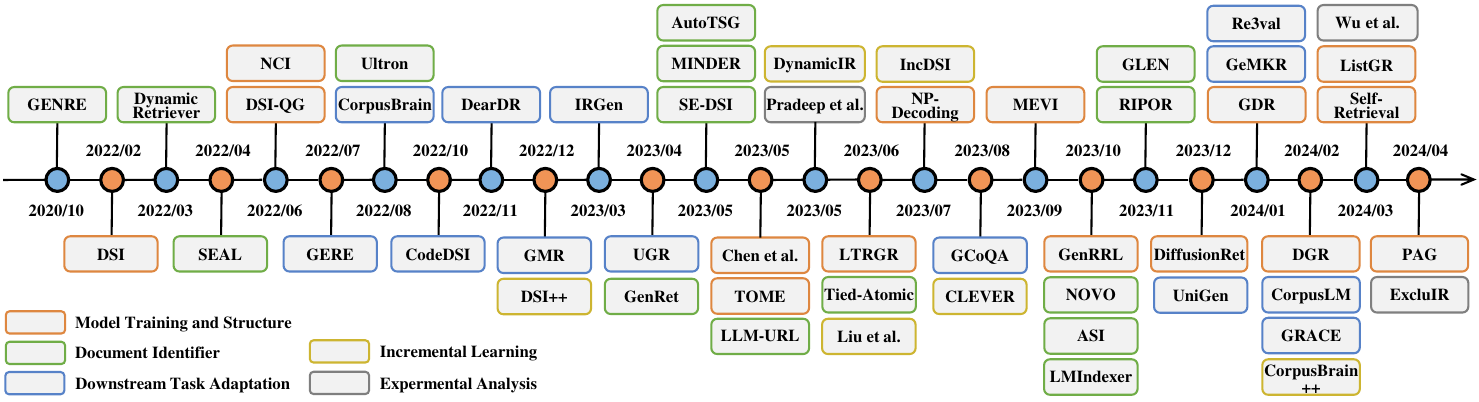}
    \vspace{-2mm}
    \caption{
    Timeline of research in generative retrieval: focus on model training and structure, document identifier design, incremental learning and downstream task adaptation.
    }
    \label{fig:gr_timeline}
\end{figure*}

In recent advancements in AIGC, generative retrieval (GR) has emerged as an promising approach in the field of information retrieval, garnering increasing interest from the academic community. Figure~\ref{fig:gr_timeline} showcases a timeline of the GR methods. Initially, GENRE~\cite{genre} proposed to retrieve entities by generating their unique names through constrained beam search via a pre-built entity prefix tree, achieving advanced entity retrieval performance. Subsequently, Metzler et al.~\cite{metzler2021rethinking} envisioned a model-based information retrieval framework aiming to combine the strengths of traditional document retrieval systems and pre-trained language models to create systems capable of providing expert-quality answers in various domains.

\begin{figure*}[!t]
    \centering
    \includegraphics[width=0.9\linewidth]{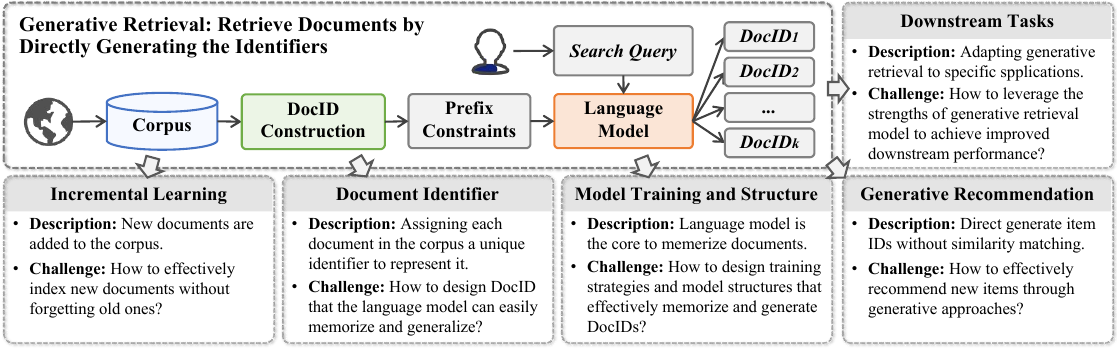}
    \vspace{-2mm}
    \caption{
    A conceptual framework for a generative retrieval system, with a focus on challenges in incremental learning, identifier construction, model training and structure, and integration with downstream tasks and recommendation systems.
    }
    \label{fig:gr_modules}
\end{figure*}

Following their lead, a diverse range of methods including DSI~\cite{dsi}, DynamicRetriever~\cite{dynamicretriever}, SEAL~\cite{seal}, NCI~\cite{nci}, etc., have been developed, with a continuously growing body of work. These methods explore various aspects such as model training and architectures, document identifiers, incremental learning, task-specific adaptation, and generative recommendations. Figure~\ref{fig:gr_modules} presents an overview of the GR system and we'll provide an in-depth discussion of each associated challenge in the following sections.

\subsection{Model Training and Structure}
\label{sec:model_training_and_structure}
One of the core components of GR is the model training and structure, aiming to enhance the model's ability to memorize documents in the corpus.

\subsubsection{Model Training}
\label{sec:gr_training}
To effectively train generative models for indexing documents, the standard approach is to train the mapping from queries to relevant DocIDs, based on standard sequence-to-sequence (seq2seq) training methods, as described in Equation~\eqref{eq:background_gr}. This method has been widely used in numerous GR research works, such as DSI~\cite{dsi}, NCI~\cite{nci}, SEAL~\cite{seal}, etc. 
Moreover, a series of works have proposed various model training methods tailored for GR tasks to further enhance GR retrieval performance, such as sampling documents or generating queries based on document content to serve as pseudo queries for data augmentation; or including training objectives for document ranking. 

Specifically, DSI~\cite{dsi} proposed two training strategies: one is “indexing”, that is, training the model to associate document tokens with their corresponding DocIDs, where DocIDs are pre-built based on documents in corpus, which will be discussed in detail in Section~\ref{sec:docid_design}; the other is “retrieval”, using labeled query-DocID pairs to fine-tune the model. Notably, DSI was the first to realize a differentiable search index based on the Transformer~\cite{transformer} structure, showing good performance in web search~\cite{ms_marco} and question answering~\cite{nq} scenarios. 
Next, a series of methods propose training methods for data augmentation and improving GR model ranking ability

\textbf{Sampling Document Pieces as Pseudo Queries.} In the same era, DynamicRetriever~\cite{dynamicretriever}, also based on the encoder-decoder model, constructed a model-based IR system by initializing the encoder with a pre-trained BERT~\cite{bert}. Besides, DynamicRetriever utilizes passages, sampled terms and N-grams to serve as pseudo queries to enhance the model's memorization of DocIDs.
Formally, the training methods can be summarized as follows:
\begin{align}
\text{Sampled Document}&: d_{s_i} \longrightarrow \text{DocID}, i \in \{1,...,k_{d_s}\}, \label{eq:sample_doc}\\
\text{Labeled Query}&: q_i \longrightarrow \text{DocID}, i \in \{1,...,k_q\}, \label{eq:label_query}
\end{align}
where $d_{s_i}$ and $q_i$ denote each of the $k_{d_s}$ sampled document text and each of the $k_q$ labeled query for the corresponding DocID, respectively.

\textbf{Generating Pseudo Queries from Documents.} Following DSI, the NCI~\cite{nci} model was trained using a combination of labeled query-document pairs and augmented pseudo query-document pairs. Specifically, NCI proposed two strategies: one using the DocT5Query~\cite{doct5query} model as a query generator, generating pseudo queries for each document in the corpus through beam search; the other strategy directly uses the document as a query, as stated in Equation~\eqref{eq:sample_doc}. Similarly, DSI-QG~\cite{dsiqg} also proposed using a query generator to enhance training data, establishing a bridge between indexing and retrieval in DSI. This data augmentation method has been proven in subsequent works to be an effective method to enhance the model's memorization for DocIDs, which can be expressed as follows:
\begin{equation}
\text{Pseudo Query}: q_{s_i} \longrightarrow \text{DocID}, i \in \{1,...,k_{q_s}\}, \label{eq:pseudo_query}
\end{equation}
where $q_{s_i}$ represents each of the $k_{q_s}$ generated pseudo query for the corresponding DocID.

\textbf{Improving Ranking Capability.} Additionally, a series of methods focus on further optimizing the ranking capability of GR models. Chen et al.~\cite{understand-dsi} proposed a multi-task distillation method to improve retrieval quality without changing the model structure, thereby obtaining better indexing and ranking capabilities. Meanwhile, LTRGR~\cite{ltrgr} introduced a ranking loss to train the model in ranking paragraphs.
Subsequently,~\cite{genrrl} proposed GenRRL, which improves ranking quality through reinforcement learning with relevance feedback, aligning token-level DocID generation with document-level relevance estimation. 
Moreover, ~\cite{dgr} introduced DGR, which enhances generative retrieval through knowledge distillation. Specifically, DGR uses a cross-encoder as a teacher model, providing fine-grained passage ranking supervision signals, and then optimizes the model with a distilled RankNet loss. ListGR~\cite{listgr} defined positional conditional probabilities, emphasizing the importance of the generation order of each DocID in the list. In addition, ListGR employs relevance calibration that adjusts the generated list of DocIDs to better align with the labeled ranking list.
See Table~\ref{tab:gr_works} for a detailed comparison of GR methods.

\subsubsection{Model Structure}
\label{sec:gr_structure}
Basic generative retrieval models mostly use pre-trained encoder-decoder structured generative models, such as T5~\cite{t5} and BART~\cite{bart}, fine-tuned for the DocID generation task. To better adapt to the GR task, researchers have proposed a series of specifically designed model structures~\cite{nci, tome, np-decoding, mevi, diffusion-ret, gdr, self-retrieval}.

\textbf{Model Decoding Methods.} For the semantic structured DocID proposed by DSI~\cite{dsi}, NCI~\cite{nci} designed a Prefix-Aware Weight-Adaptive (PAWA) decoder. By adjusting the weights at different positions of DocIDs, this decoder can capture the semantic hierarchy of DocIDs.
To allow the GR model to utilize both own parametric knowledge and external information, NP-Decoding~\cite{np-decoding} proposed using non-parametric contextualized word embeddings (as external memory) instead of traditional word embeddings as the input to the decoder.
Additionally, PAG~\cite{PAG} proposed a planning-ahead generation approach, which first decodes the set-based DocID to approximate document-level scores, and then continues to decode the sequence-based DocID on this basis. 

\textbf{Combining Generative and Dense Retrieval Methods.} Combining seq2seq generative models with dual-encoder retrieval models, MEVI~\cite{mevi} utilizes Residual Quantization (RQ)~\cite{RQ} to organize documents into hierarchical clusters, enabling efficient retrieval of candidate clusters and precise document retrieval within those clusters. Similarly, Generative Dense Retrieval (GDR)~\cite{gdr} proposed to first broadly match queries to document clusters, optimizing for interaction depth and memory efficiency, and then perform precise, cluster-specific document retrieval, boosting both recall and scalability.

\textbf{Utilizing Multiple Models.} TOME~\cite{tome} proposed to decompose the GR task into two stages, first generating text paragraphs related to the query through an additional model, then using the GR model to generate the URL related to the paragraph. DiffusionRet~\cite{diffusion-ret} proposed to first use a diffusion model (SeqDiffuSeq~\cite{SeqDiffuSeq}) to generate a pseudo-document from a query, where the generated pseudo-document is similar to real documents in length, format, and content, rich in semantic information; Then, it employs another generative model to perform retrieval based on N-grams, similar to the process used by SEAL\cite{seal}, leveraging an FM-Index\cite{fm_index} for generating N-grams found in the corpus.
Self-Retrieval~\cite{self-retrieval} fully integrated indexing, retrieval, and evaluation into a single large language model. It generates natural language indices and document segments, and performs self-evaluation to score and rank the generated documents.

\subsection{Design of Document Identifiers}
\label{sec:docid_design}
Another essential component of generative retrieval is document representation, also known as document identifiers (DocIDs), which act as the target outputs for the GR model. Accurate document representations are crucial as they enable the model to more effectively memorize document information, leading to enhanced retrieval performance. Table~\ref{tab:gr_works} provides a detailed comparison of the states, data types, and order of DocIDs across numerous GR methods.
In the following sections, we will explore the design of DocIDs from two categories: numeric-based identifiers and text-based identifiers.

\subsubsection{Numeric-based Identifiers}
\label{sec:numeric_docid}
An intuitive method to represent documents is by using a single number or a series of numbers, referred to as DocIDs. Existing methods have designed both static and learnable DocIDs.

\textbf{Static DocIDs.} 
Initially, DSI~\cite{dsi} introduced three numeric DocIDs to represent documents: (1) Unstructured Atomic DocID: a unique integer identifier is randomly assigned to each document, containing no structure or semantic information. (2) Naively Structured String DocID: treating random integers as divisible strings, implementing character-level DocID decoding to replace large softmax output layers. (3) Semantically Structured DocID: introducing semantic structure through hierarchical $k$-means method, allowing semantically similar documents to share prefixes in their identifiers, effectively reducing the search space. Concurrently, DynamicRetriever~\cite{dynamicretriever} also built a model-based IR system based on unstructured atomic DocID.
Subsequently, Ultron~\cite{ultron} encoded documents into a latent semantic space using BERT~\cite{bert}, and compressed vectors into a smaller semantic space via Product Quantization (PQ)~\cite{PQ, Optimized_PQ}, preserving semantic information. Each document's PQ code serves as its semantic identifier. MEVI~\cite{mevi} clusters documents using Residual Quantization (RQ)~\cite{RQ} and utilizes dual-tower and seq2seq model embeddings for a balanced performance in large-scale document retrieval.

\textbf{Learnable DocIDs.} 
Unlike previous static DocIDs, GenRet~\cite{genret} proposed learnable document representations, transforming documents into DocIDs through an encoder, then reconstructs documents from DocIDs using a decoder, trained to minimize reconstruction error. Furthermore, it used progressive training and diversity clustering for optimization. To ensure that DocID embeddings can reflect document content, Tied-Atomic~\cite{gen_as_dense} proposed to link document text with token embeddings and employs contrastive loss for DocID generation.
LMIndexer~\cite{lmindexer} and ASI~\cite{asi} learned optimal DocIDs through semantic indexing, with LMIndexer using a reparameterization mechanism for unified optimization, facilitating efficient retrieval by aligning semantically similar documents under common DocIDs. ASI extends this by establishing an end-to-end retrieval framework, incorporating semantic loss functions and reparameterization to enable joint training. 
Furthermore, RIPOR~\cite{ripor} treats the GR model as a dense encoder to encode document content. It then splits these representations into vectors via RQ~\cite{RQ}, creating unique DocID sequences. Furthermore, RIPOR implements a prefix-guided ranking optimization, increasing relevance scores for prefixes of pertinent DocIDs through margin decomposed pairwise loss during decoding. 

In summary, numeric-based document representations can utilize the embeddings of dense retrievers, obtaining semantically meaningful DocID sequences through methods such as $k$-means, PQ~\cite{PQ}, and RQ~\cite{RQ}; they can also combine encoder-decoder GR models with bi-encoder DR models to achieve complementary advantages~\cite{gen_as_dense, mevi}.

\definecolor{lightgray}{gray}{0.94}

\begin{table*}[!t]
\centering
\caption{Comparisons of representative generative retrieval methods, focusing on document identifier, training data augmentation, and training objective.}
\vspace{-2mm}
\label{tab:gr_works}
\setlength\tabcolsep{4.1pt}
\scriptsize
\renewcommand{\arraystretch}{1.06} % 调整行高
\begin{tabular}{lcccccccc}
\toprule
\multirow{2}[2]{*}{\textbf{Method}} & \multicolumn{3}{c}{\textbf{Document Identifier}} & \multicolumn{2}{c}{\textbf{Training Data Augmentation}} & \multicolumn{3}{c}{\textbf{Training Objective}} \\ 
\cmidrule(lr){2-4} \cmidrule(lr){5-6} \cmidrule(lr){7-9}
& \textbf{State} & \textbf{Data Type} & \textbf{Order} & \textbf{Sample Doc} & \textbf{Doc2Query} & \textbf{Seq2seq} & \textbf{DocID} & \textbf{Ranking} \\ 
\midrule
\rowcolor{lightgray}GENRE~\cite{genre} & Static & Text & Sequence & - & - & \ding{51} & - & - \\
DSI~\cite{dsi} & Static & Numeric & Sequence & \ding{51} & - & \ding{51} & - & - \\
\rowcolor{lightgray}DynamicRetriever~\cite{dynamicretriever} & Static & Numeric & Sequence & \ding{51} & - & \ding{51} & - & - \\
SEAL~\cite{seal} & Static & Text & Sequence & \ding{51} & - & \ding{51} & - & - \\
\rowcolor{lightgray}DSI-QG~\cite{dsiqg} & Static & Numeric & Sequence & - & \ding{51} & \ding{51} & - & - \\
NCI~\cite{nci} & Static & Numeric & Sequence & \ding{51} & \ding{51} & \ding{51} & - & - \\
\rowcolor{lightgray}Ultron~\cite{ultron} & Static & Numeric/Text & Sequence & \ding{51} & \ding{51} & \ding{51} & - & - \\
CorpusBrain~\cite{corpusbrain} & Static & Text & Sequence & \ding{51} & - & \ding{51} & - & - \\
\rowcolor{lightgray}GenRet~\cite{genret} & Learnable & Numeric & Sequence & - & \ding{51} & \ding{51} & \ding{51} & - \\
AutoTSG~\cite{autotsg} & Static & Text & Set & - & \ding{51} & \ding{51} & - & - \\
\rowcolor{lightgray}SE-DSI~\cite{se-dsi} & Static & Text & Sequence & \ding{51} & - & \ding{51} & - & - \\
Chen et al.~\cite{understand-dsi} & Static & Numeric & Sequence & \ding{51} & \ding{51} & \ding{51} & - & \ding{51} \\
\rowcolor{lightgray}LLM-URL~\cite{llm-url} & Static & Text & Sequence & - & - & - & - & - \\
MINDER~\cite{minder} & Static & Text & Sequence & - & \ding{51} & \ding{51} & - & - \\
\rowcolor{lightgray}LTRGR~\cite{ltrgr} & Static & Text & Sequence & - & \ding{51} & \ding{51} & - & \ding{51} \\
NOVO~\cite{novo} & Learnable & Text & Set & \ding{51} & - & - & \ding{51} & - \\
\rowcolor{lightgray}GenRRL~\cite{genrrl} & Static & Text & Sequence & - & \ding{51} & \ding{51} & - & \ding{51} \\
LMIndexer~\cite{lmindexer} & Learnable & Numeric & Sequence & - & \ding{51} & \ding{51} & \ding{51} & - \\
\rowcolor{lightgray}ASI~\cite{asi} & Learnable & Numeric & Sequence & - & \ding{51} & \ding{51} & \ding{51} & - \\
RIPOR~\cite{ripor} & Learnable & Numeric & Sequence & - & \ding{51} & \ding{51} & \ding{51} & \ding{51} \\
\rowcolor{lightgray}GLEN~\cite{glen} & Learnable & Text & Sequence & - & \ding{51} & \ding{51} & \ding{51} & \ding{51} \\
DGR~\cite{dgr} & Static & Text & Sequence & - & \ding{51} & \ding{51} & - & \ding{51} \\
\rowcolor{lightgray}ListGR~\cite{listgr} & Static & Numeric & Sequence & - & \ding{51} & \ding{51} & - & \ding{51} \\
% PAG~\cite{PAG} & Static & Numeric & Set/Sequence & - & - & \ding{51} & - & - \\
\bottomrule
\end{tabular}
\end{table*}

\subsubsection{Text-based Identifiers}
\label{sec:text_docid}
Text-based DocIDs have the inherent advantage of effectively leveraging the strong capabilities of pre-trained language models and offering better interpretability.

\textbf{Document Titles.} 
The most straightforward text-based identifier is the document title, which requires each title to uniquely represent a document in the corpus, otherwise, it would not be possible to accurately retrieve a specific document. The Wikipedia corpus used in the KILT~\cite{kilt} benchmark, due to its well-regulated manual annotation, has a unique title corresponding to each document. Thus, GENRE~\cite{genre}, based on the title as DocID and leveraging the generative model BART~\cite{bart} and pre-built DocID prefix, achieved superior retrieval performance across 11 datasets in KILT. Following GENRE, GERE~\cite{gere}, CorpusBrain~\cite{corpusbrain}, Re3val~\cite{re3val}, and CorpusBrain++~\cite{corpusbrain++} also based their work on title DocIDs for Wikipedia-based tasks. Notably, LLM-URL~\cite{llm-url} directly generated URLs using ChatGPT prompts, achieving commendable performance after removing invalid URLs.
However, in the web search scenario~\cite{ms_marco}, document titles in the corpus often have significant duplication and many meaningless titles, making it unfeasible to use titles alone as DocIDs. Thus, Ultron~\cite{ultron} effectively addressed this issue by combining URLs and titles as DocIDs, identifying documents through keywords in web page URLs and titles.

\textbf{Sub-strings of Documents.} 
To increase the flexibility of DocIDs, SEAL~\cite{seal} proposed a sub-string identifier, representing documents with any N-grams within them. Using FM-Index (a compressed full-text sub-string index)~\cite{fm_index}, SEAL could generate N-grams present in the corpus to retrieve all documents containing those N-grams, scoring and ranking documents based on the frequency of N-grams in each document and the importance of N-grams. Following SEAL, various GR models~\cite{ugr, minder, ltrgr, dgr} also utilized sub-string DocIDs and FM-Index during inference. For a more comprehensive representation of documents, MINDER~\cite{minder} proposed multi-view identifiers, including generated pseudo queries from document content via DocT5Query~\cite{doct5query}, titles, and sub-strings. This multi-view DocID was also used in LTRGR~\cite{ltrgr} and DGR~\cite{dgr}. 

\textbf{Term Sets.} 
Unlike the sequential DocIDs described earlier, AutoTSG~\cite{autotsg} proposed a term set-based document representation, using keywords extracted from titles and content, rather than predefined sequences, allowing for retrieval of the target document as long as the generated term set is included in the extracted keywords. Recently, PAG~\cite{PAG} also constructed DocIDs based on sets of key terms, disregarding the order of terms, which is utilized for approximating document relevance in decoding.

\textbf{Learnable DocIDs.}
Text-based identifiers can also be learnable. Similarly based on term-sets, NOVO~\cite{novo} proposed learnable continuous N-grams constituting term-set DocIDs. Through denoising query modeling, the model learned to generate queries from documents with noise, thereby implicitly learning to filter out document N-grams more relevant to queries. NOVO also improves the document's semantic representation by updating N-gram embeddings.
Later, GLEN\cite{glen} uses dynamic lexical DocIDs and follows a two-phase index learning strategy. First, it assigns DocIDs by extracting keywords from documents using self-supervised signals. Then, it refines DocIDs by integrating query-document relevance through two loss functions. During inference, GLEN ranks documents using DocID weights without additional overhead.

\subsection{Incremental Learning on Dynamic Corpora}
\label{sec:continual_learning}
Prior studies have focused on generative retrieval from static document corpora. However, in reality, the documents available for retrieval are continuously updated and expanded. To address this challenge, researchers have developed a range of methods to optimize GR models for adapting to dynamic corpora.

\textbf{Optimizer and Document Rehearsal.}
At first, DSI++~\cite{dsi++} aims to address the incremental learning challenges encountered by DSI~\cite{dsi}. DSI++ modifies the training by optimizing flat loss basins through the Sharpness-Aware Minimization (SAM) optimizer, stabilizing the learning process of the model. It also employs DocT5Query~\cite{doct5query} to generate pseudo queries for documents in the existing corpus as training data augmentation, mitigating the forgetting issue of GR models.

\textbf{Constrained Optimization}
Addressing the scenario of real-time addition of new documents, such as news or scientific literature IR systems, IncDSI~\cite{incdsi} views the addition of new documents as a constrained optimization problem to find optimal representations for the new documents. This approach aims to (1) ensure new documents can be correctly retrieved by their relevant queries, and (2) maintain the retrieval performance of existing documents unaffected. %IncDSI manages to add each document within approximately 20-50ms, significantly reducing the time and computational resources required compared to full model retraining, while maintaining competitive retrieval performance.

\textbf{Incremental Product Quantization.}
CLEVER~\cite{clever}, based on Product Quantization (PQ)~\cite{PQ}, proposes Incremental Product Quantization (IPQ) for generating PQ codes as DocIDs for documents. Compared to traditional PQ methods, IPQ designs two adaptive thresholds to update only a subset of centroids instead of all, maintaining the indices of updated centroids constant. This method reduces computational costs and allows the system to adapt flexibly to new documents. %To mitigate forgetting, CLEVER employs a memory-enhanced learning mechanism, maintaining a dynamic memory bank to store example documents similar to the new ones.

\textbf{Fine-tuning Adatpers for Specific Tasks.}
CorpusBrain++~\cite{corpusbrain++} introduces the KILT++ benchmark for continuously updated KILT~\cite{kilt} tasks and designs a dynamic architecture paradigm with a backbone-adapter structure. By fixing a shared backbone model to provide basic retrieval capabilities while introducing task-specific adapters to incrementally learn new documents for each task, it effectively avoids catastrophic forgetting. During training, CorpusBrain++ generates pseudo queries for new document sets and continues to pre-train adapters for specific tasks. %Moreover, it employs document clustering based on semantic similarity and a retraining strategy to maintain memory of older documents by revisiting them.

\subsection{Downstream Task Adaption}
\label{sec:downstream_adaption}
Generative retrieval methods, apart from addressing retrieval tasks individually, have been tailored to various downstream generative tasks. These include fact verification~\cite{fever}, entity linking~\cite{aida}, open-domain QA~\cite{nq}, dialogue~\cite{wow}, slot filling~\cite{zsre}, among others, as well as knowledge-intensive tasks~\cite{kilt}, code~\cite{CodeXGLUE}, conversational QA~\cite{TopiOCQA}, and multi-modal retrieval scenarios~\cite{MS_COCO}, demonstrating superior performance and efficiency. These methods are discussed in terms of separate training, joint training, and multi-modal generative retrieval.

\subsubsection{Separate Training}
\label{sec:gr_seperate_train}
For fact verification tasks~\cite{fever}, which involve determining the correctness of input claims, GERE~\cite{gere} proposed using an encoder-decoder-based GR model to replace traditional indexing-based methods. Specifically, GERE first utilizes a claim encoder to encode input claims, and then generates document titles related to the claim through a title decoder to obtain candidate sentences for corresponding documents. %Finally, an evidence decoder generates evidence sentence identifiers, resulting in improvements in time, memory consumption, and performance.

\textbf{Knowledge-Intensive Language Tasks.}
For Knowledge-Intensive Language Tasks (KILT)~\cite{kilt}, CorpusBrain~\cite{corpusbrain} introduced three pre-training tasks to enhance the model's understanding of query-document relationships at various granularities: Internal Sentence Selection, Leading Paragraph Selection, and Hyperlink Identifier Prediction.
Similarly, UGR~\cite{ugr} proposed using different granularities of N-gram DocIDs to adapt to various downstream tasks, unifying different retrieval tasks into a single generative form. UGR achieves this by letting the GR model learn prompts specific to tasks, generating corresponding document, passage, sentence, or entity identifiers.

Futhermore, DearDR~\cite{deardr} utilizes remote supervision and self-supervised learning techniques, using Wikipedia page titles and hyperlinks as training data. The model samples sentences from Wikipedia documents as input and trains a self-regressive model to decode page titles or hyperlinks, or both, without the need for manually labeled data. %The model achieves zero-shot implementation of Wikipedia-based fact verification tasks and further optimizes performance through fine-tuning.
Re3val~\cite{re3val} proposes a retrieval framework combining generative reordering and reinforcement learning. It first reorders retrieved page titles using context information obtained from a dense retriever, then optimizes the reordering using the REINFORCE algorithm to maximize rewards generated by constrained decoding. %By improving page title reordering and context selection, Re3val achieves more accurate information retrieval.

\textbf{Multi-hop retrieval.}
In multi-hop retrieval tasks, which require iterative document retrievals to gather adequate evidence for answering a query, GMR~\cite{gmr} proposed to employ language model memory and multi-hop memory to train a generative retrieval model, enabling it to memorize the target corpus and simulate real retrieval scenarios through constructing pseudo multi-hop query data, achieving dynamic stopping and efficient performance in multi-hop retrieval tasks.

\textbf{Code Retrieval.}
CodeDSI~\cite{codedsi} is an end-to-end generative code search method that directly maps queries to pre-stored code samples' DocIDs instead of generating new code. Similar to DSI~\cite{dsi}, it includes indexing and retrieval stages, learning to map code samples and real queries to their respective DocIDs. CodeDSI explores different DocID representation strategies, including direct and clustered representation, as well as numerical and character representations. %showing superior performance compared to traditional methods on 1K and 10K scales, with numerical DocIDs performing better than alphabetic ones.

\textbf{Conversational Question Answering.}
GCoQA~\cite{GCoQA} is a generative retrieval method for conversational QA systems that directly generates DocIDs for passage retrieval. This method focuses on key information in the dialogue context at each decoding step, achieving more precise and efficient passage retrieval and answer generation, thereby improving retrieval performance and overall system efficiency.

\subsubsection{Joint Training}
\label{sec:gr_joint_train}
The methods in the previous section involve separately training generative retrievers and downstream task generators. However, due to the inherent nature of GR models as generative models, a natural advantage lies in their ability to be jointly trained with downstream generators to obtain a unified model for retrieval and generation tasks. 
% This section will introduce methods for joint training.

\textbf{Multi-decoder Structure.}
UniGen~\cite{unigen} proposes a unified generation framework to integrate retrieval and question answering tasks, bridging the gap between query input and generation targets using connectors generated by large language models. UniGen employs shared encoders and task-specific decoders for retrieval and question answering, introducing iterative enhancement strategies to continuously improve the performance of both tasks. %It demonstrates superior performance on both web search and question answering tasks.

\textbf{Multi-task Training.}
Later, CorpusLM~\cite{corpuslm} introduces a unified language model that integrates GR, closed-book generation, and retrieval-augmented generation to handle various knowledge-intensive tasks. The model adopts a multi-task learning approach and introduces ranking-guided DocID decoding strategies and continuous generation strategies to improve retrieval and generation performance. In addition, CorpusLM designs a series of auxiliary DocID understanding tasks to deepen the model's understanding of DocID semantics. %Experimental results validate the effectiveness and potential of CorpusLM (T5~\cite{t5} and Llama2~\cite{llama2} variants) for knowledge-intensive language tasks.

\subsubsection{Multi-modal Generative Retrieval}
\label{sec:multi_modal}
Generative retrieval methods can also leverage multi-modal data such as text, images, etc., to achieve end-to-end multi-modal retrieval.

\textbf{Tokenizing Images to DocID Sequences.}
At first, IRGen~\cite{irgen} transforms image retrieval problems into generative problems, predicting relevant discrete visual tokens, i.e., image identifiers, through a seq2seq model given a query image. IRGen proposed a semantic image tokenizer, which converts global image features into short sequences capturing high-level semantic information. %Unlike traditional methods that handle feature extraction and ANN search separately, IRGen achieves end-to-end differentiable search, optimizing directly from the final retrieval target, thereby enhancing retrieval accuracy and efficiency.

\textbf{Advanced Model Training and Structure.}
Later, GeMKR~\cite{gemkr} combines LLMs' generation capabilities with visual-text features, designing a generative knowledge retrieval framework. It first guides multi-granularity visual learning using object-aware prefix tuning techniques to align visual features with LLMs' text feature space, achieving cross-modal interaction. GeMKR then employs a two-step retrieval process: generating knowledge clues closely related to the query and then retrieving corresponding documents based on these clues. %It aims to improve knowledge retrieval efficiency and accuracy in multi-modal scenarios.
GRACE~\cite{gemkr} achieves generative cross-modal retrieval method by assigning unique identifier strings to images and training multi-modal large language models (MLLMs)~\cite{OpenFlamingo} to memorize the association between images and their identifiers. The training process includes (1) learning to memorize images and their corresponding identifiers, and (2) learning to generate the target image identifiers from textual queries. GRACE explores various types of image identifiers, including strings, numbers, semantic and atomic identifiers, to adapt to different memory and retrieval requirements.

\subsubsection{Generative Recommender Systems}
Recommendation systems, as an integral part of the information retrieval, are currently undergoing a paradigm shift from discriminative models to generative models. Generative recommendation systems do not require the computation of ranking scores for each item followed by database indexing, but instead accomplish item recommendations through the direct generation of IDs. In this section, several seminal works, including P5~\cite{p5}, GPT4Rec~\cite{GPT4Rec}, TIGER~\cite{TIGER}, SEATER~\cite{seater}, IDGenRec~\cite{IDGenRec}, LC-Rec~\cite{LCRec} and ColaRec~\cite{ColaRec}, are summarized to outline the development trends in generative recommendations.

P5~\cite{p5} transforms various recommendation tasks into different natural language sequences, designing a universal, shared framework for recommendation completion. This method, by setting unique training objectives, prompts, and prediction paradigms for each recommendation domain's downstream tasks, serves well as a backbone model, accomplishing various recommendation tasks through generated text. %This approach demonstrates the viability and flexibility of generative models in recommendation systems.
In generative retrieval, effective indexing identifiers have been proven to significantly enhance the performance of generative methods. Similarly, TIGER~\cite{TIGER} initially learns a residual quantized autoencoder to generate semantically informative indexing identifiers for different items. It then trains a transformer-based encoder-decoder model with this semantically informative indexing identifier sequence to generate item identifiers for recommending the next item based on historical sequences.

Focusing solely on semantic information and overlooking the collaborative filtering information under the recommendation context might limit the further development of generative models. Therefore, after generating semantic indexing identifiers similar to TIGER using a residual quantized autoencoder with uniform semantic mapping, LC-Rec~\cite{LCRec} also engages in a series of alignment tasks, including sequential item prediction, explicit index-language alignment, and recommendation-oriented implicit alignment. Based on the learned item identifiers, it integrates semantic and collaborative information, enabling large language models to better adapt to sequence recommendation tasks.

IDGenRec~\cite{IDGenRec} innovatively combines generative recommendation systems with large language models by using human language tokens to generate unique, concise, semantically rich and platform-agnostic texual identifiers for recommended items. The framework includes a text ID generator trained on item metadata with a diversified ID generation algorithm, and an alternating training strategy that optimizes both the ID generator and the LLM-based recommendation model for improved performance and accuracy in sequential recommendations. %The zero-shot performance of IDGenRec is comparable to, or even surpasses, certain traditional recommendation models that rely on supervised training, highlighting its potential as a foundational model for recommendation systems.
SEATER~\cite{seater} designs a balanced k-ary tree-structured indexes, using a constrained k-means clustering method to recursively cluster vectors encoded from item texts, obtaining equal-length identifiers. Compared to the method proposed by DSI~\cite{dsi}, this balanced k-ary tree index maintains semantic consistency at every level. It then trains a Transformer-based encoder-decoder model and enhances the semantics of each level of indexing through contrastive learning and multi-task learning.
ColaRec~\cite{ColaRec} integrates collaborative filtering signals and content information by deriving generative item identifiers from a pretrained recommendation model and representing users via aggregated item content. Then it uses an item indexing generation loss and contrastive loss to align content-based semantic spaces with collaborative interaction spaces, enhancing the model's ability to recommend items in an end-to-end framework.

\label{sec:recommender}

\section{Reliable Response Generation: Direct Information Accessing with Generative Language Models}
\label{sec:response_generation}

\begin{figure*}[!t]
    \centering
    \includegraphics[width=0.85\linewidth]{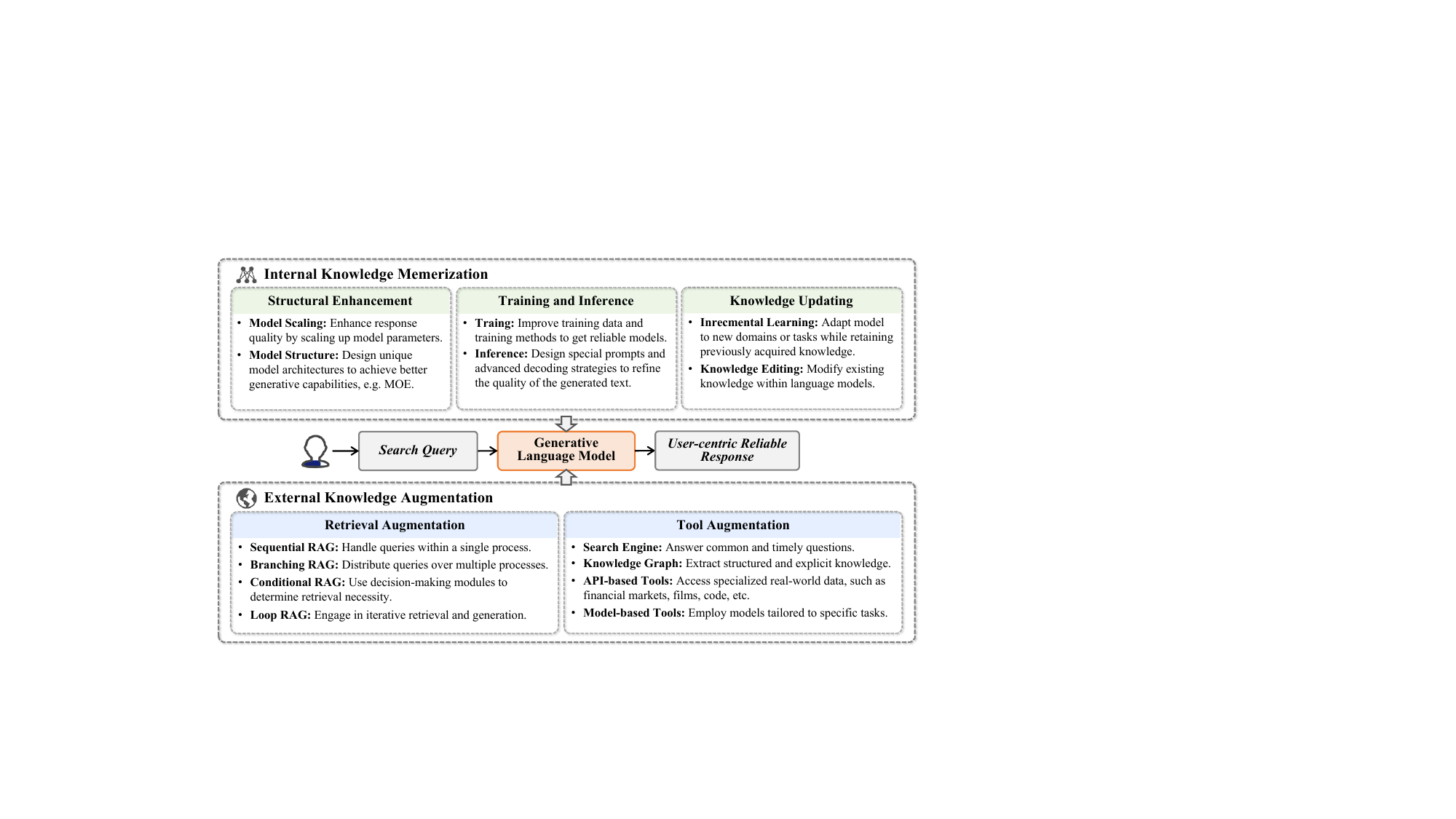}
    \vspace{-2mm}
    \caption{
    An Illustration of strategies for enhancing language models to generate user-centric and reliable responses, including model internal knowledge memorization and external knowledge augmentation.
    }
    \label{fig:gen_modules}
\end{figure*}

The rapid advancement of large language models has positioned them as a novel form of IR system, capable of generating reliable responses directly aligned with users' informational needs. This not only saves the time users would otherwise spend on collecting and integrating information but also provides personalized, user-centric answers tailored to individual users.

However, challenges remain in creating a grounded system that delivers faithful answers, such as hallucination, prolonged inference time, and high operational costs. 
This section will outline strategies for constructing a faithful GenIR system by: (1) Optimizing the GenIR model internally, (2) Enhancing the model with external knowledge, (3) Increasing accountability, and (4) Developing personalized information assistants.

\subsection{Internal Knowledge Memorization}
\label{sec:model_internal_enhance}

To develop a user-friendly and reliable IR system, the generative model should be equipped with comprehensive internal knowledge. Optimization of the backbone generative model can be categorized into three aspects: structural enhancements, training strategies, and inference techniques. 
The overview of this section is shown in the green part of Figure~\ref{fig:gen_modules}.

\subsubsection{Model Structure}
\label{sec:structural_enhance}

With the advent of generative models, various methods have been introduced to improve model structure and enhance generative reliability. We aim to discuss the crucial technologies contributing to this advancement in this subsection.

\textbf{(1) Model Scaling} Model parameter scaling is a pivotal factor influencing performance. Contemporary language models predominantly employ the Transformer architecture, and scaling both the model parameters and the training data enhances the model's capacity to retain knowledge and capabilities~\cite{scalinglaw}. For instance, in the GPT~\cite{gpt,gpt2,gpt3,gpt4} series and LLaMA~\cite{llama,llama2} family, larger models tend to perform better on diverse downstream tasks, including few-shot learning, language understanding, and generation~\cite{palm}. Additionally, scaling the model contributes to improved instruction-following capabilities~\cite{infobench}, enabling a more adept comprehension of user intent and generating responses that better satisfy user requests.

\textbf{(2) Model Integration} Model integration is an effective method to enhance the reliability of generated outputs by capitalizing on the diverse strengths of various models. The predominant approach is the Mixture of Experts (MoE)~\cite{adaptive1991}, which utilizes a gating mechanism to selectively activate sections of network parameters during inference, greatly increasing the effective parameters without inflating inference costs~\cite{switchtransformers,gshard,glam,mixtral8x7b}. This method also boasts impressive scalability, with efficacy augmented alongside the expanding parameter volume and the number of expert models~\cite{moescalinglaw}. Alternatively, the LLM-Blender framework~\cite{llm-blender} employs a ranker and a fuser to combine answers from various LLMs, including black-box models, but faces high deployment costs.

\subsubsection{Training and Inference}
\label{sec:training_inference_enhance}

In the model training stage, methods to enhance the reliability of answers can be categorized into two aspects: training data optimization and training methods optimization.

\textbf{(1) Training Data Optimization} The quality of training data substantially affects the reliability of model outputs. Noise, misinformation, and incomplete information can disrupt the learning process, leading to hallucinations and other issues. To address this, \cite{textbooks} used GPT-3.5 to artificially create textbooks filled with examples and language descriptions as training data, resulting in significant improvements on downstream tasks after minor fine-tuning. LIMA~\cite{LIMA} used dialogues from community forums to construct a small-scale fine-tuning dataset, enhancing the model's conversation capabilities during the alignment phase. To reduce redundancies in crawled internet data, Lee et al.~\cite{Lee_dedup_training_data} combined suffix arrays~\cite{suffixarray} and MinHash~\cite{minhash} to approximate matching and deduplicate the training dataset, reducing direct reproduction from the same source.

\textbf{(2) Training Methods Optimization} Beyond conventional training methods, additional techniques have been proposed to improve the factuality of model outputs. MixCL~\cite{mixcl} incorporates contrastive learning into the training objective, using an external knowledge base to identify correct snippets and reduce the probability of generating incorrect tokens, thus enhancing model reliability. CaliNet~\cite{CaliNet} utilizes a contrastive method to assess erroneous knowledge learned by the model and fine-tunes the parameters of the FFN layer to rectify these errors. FactTune~\cite{FactTune} incorporates factuality assessment during the RLHF phase, using automatic evaluation methods like FactScore~\cite{FActScore} to rank outputs and employing DPO~\cite{dpo} to teach the model factuality preference.

Apart from enhancing the internal knowledge reliability during training, the inference stage significantly impacts the reliability of answers. The overall inference process consists of user input and the model's token decoding, and approaches to increase generation reliability can be divided into prompt engineering and decoding strategy.

\textbf{(3) Prompt Engineering} Prompting methods play a vital role in guiding the model. A well-designed prompt can better promote the model's internal capabilities to provide more accurate answers. The Chain-of-Thought (CoT)~\cite{COT} prompting method guides the model to explicitly decompose the question into a reasoning chain during decoding, improving response accuracy by grounding the final answer on accurate intermediate steps. Further, CoT-SC~\cite{cot-sc} samples multiple answers and chooses the most consistent one as the final answer. The Tree of Thoughts~\cite{TOT} expands CoT's single reasoning path to multiple paths, synthesizing their outcomes to arrive at the final answer. The Chain-of-Verification (CoVE)~\cite{Chainofverification} introduces a self-reflection mechanism where the LLM generates a draft response, then validates each statement for factual inaccuracies, correcting errors to enhance factual accuracy. Additionally, methods like RECITE~\cite{recite} and GenRead~\cite{genread} prompt the model to output relevant internal knowledge fragments, which are then used to bolster the question-answering process.

\textbf{(4) Decoding Strategy} Decoding strategies are another critical factor influencing the reliability of model-generated responses. An appropriate decoding method can maintain the reliability and diversity of a model's response. Nucleus Sampling~\cite{nucleus_sampling} samples within a set probability range for tokens, ensuring better diversity while balancing variety and reliability. Building on this, Factual-Nucleus Sampling~\cite{factuality_nuc} employs a dynamic, decaying threshold for token sampling, ensuring later tokens are not influenced by earlier less factual tokens. Wan et al.~\cite{faithfulness} proposed a faithfulness-aware decoding method to enhance the faithfulness of the beam-search approach by incorporating a Ranker to reorder generated sequences and a lookahead method to avoid unfaithful tokens. 

Apart from directly modifying the decoding method, several studies influence the decoding distribution by leveraging hidden layer information. DoLa~\cite{dola} uses distributional differences between hidden and output layers to prioritize newly learned factual knowledge or key terms, increasing their generation likelihood. Inference-Time Intervention (ITI)~\cite{ITI} identifies attention heads strongly correlated with response correctness, adjusts their orientations, and moderates their activation, achieving more truthful generation with minimal model interference. Shi et al.~\cite{trustingevidence} proposed CAD, comparing output distributions before and after adding extra information, reducing reliance on the model's own knowledge to avoid conflicts leading to inaccuracies.

\definecolor{lightgray}{gray}{0.94}

\begin{table*}[!t]
\centering
\caption{Comparison of representative reliable response generation methods, considering model configurations, specializations, and evaluations. For simplicity, "LM" stands for Language Modeling and "ODQA" stands for Open-Domain Question Answering. }
\vspace{-2mm}
\label{tab:gen_works}
\scriptsize
\renewcommand{\arraystretch}{1.06} % 调整行高
\begin{tabular}{
    >{\raggedright\arraybackslash}p{1.85cm} % Method
    >{\centering\arraybackslash}p{1.85cm}   % Backbone
    >{\centering\arraybackslash}p{1.1cm} % Parameters
    >{\centering\arraybackslash}p{1cm}    % Trained
    >{\centering\arraybackslash}p{1.6cm} % Specialization
    >{\centering\arraybackslash}p{4.4cm}   % Evaluation Tasks
}
\toprule
\multirow{2}[2]{*}{\textbf{Method}} & \multicolumn{3}{c}{\textbf{Model Configuration}} & \multicolumn{2}{c}{\textbf{Target Domain}} \\ 
\cmidrule(lr){2-4} \cmidrule(lr){5-6}
 & \textbf{Backbone} & \textbf{Parameters} & \textbf{Trained} & \textbf{Capability} & \textbf{Evaluation Task} \\
\midrule
\rowcolor{lightgray}GPT-3~\cite{gpt3} & Transformer & 175B & \ding{51} & General & General Tasks (LM, QA, Reasoning, ...) \\
Llama-3.1~\cite{llama3} & Transformer & 8B/70B/405B & \ding{51} & General & General Tasks \\
\rowcolor{lightgray}Mistral~\cite{mistral_7b} & Transformer & 7B/22B/123B & \ding{51} & General & General Tasks \\
PaLM~\cite{palm} & Transformer & 540B & \ding{51} & General & General Tasks \\
\rowcolor{lightgray}FactTune~\cite{FactTune} & Llama-2 & 7B & \ding{51} & Factuality & Domain-specific QA \\
GenRead~\cite{genread} & InstructGPT & 175B & $\times$ & Factuality & Knowledge-intensive Tasks \\
\rowcolor{lightgray}DoLa~\cite{dola} & LLaMA & 7B\~65B & $\times$ & Factuality & Multi-choice QA, Open-ended Generation \\
\midrule
RAG~\cite{rag} & BART & 400M & \ding{51} & Factuality & Knowledge-intensive Tasks \\
\rowcolor{lightgray}REPLUG~\cite{replug} & GPT-3 & 175B & $\times$ & Factuality & LM, Multi-choice QA, ODQA \\
FLARE~\cite{flare} & GPT-3 & 175B & $\times$ & Factuality & Knowledge-intensive Tasks \\
\rowcolor{lightgray}Self-RAG~\cite{self-rag} & Llama-2 & 7B/13B & \ding{51} & Factuality & ODQA, Reasoning, Fact Check. \\
IR-CoT~\cite{ircot} & GPT-3/Flan-T5 & 175B/11B & $\times$ & Factuality & Multi-hop QA \\
\rowcolor{lightgray}ReAct~\cite{react} & PaLM & 540B & $\times$ & Tools & Multi-hop QA, Fact Check., Decision Making \\
StructGPT~\cite{structgpt} & GPT-3/GPT-3.5 & 175B/- & $\times$ & Tools & KG-based QA, Table-based QA, Text-to-SQL \\
\rowcolor{lightgray}ToolFormer~\cite{toolformer} & GPT-J & 6B & \ding{51} & Tools & LM, Math, QA, Temporal Tasks \\
ToolLLM~\cite{ToolLLM} & LLaMA & 7B & \ding{51} & Tools & Tool Use \\
\rowcolor{lightgray}HuggingGPT~\cite{hugginggpt} & GPT-3.5 & - & $\times$ & Tools & Various Complex AI Tasks \\
\midrule
According to~\cite{According_to_Prompting} & GPT-3/Flan-T5/... & 175B/11B/... & $\times$ & Accountability & ODQA \\
\rowcolor{lightgray}IFL~\cite{IFL} & GPT-J & 6B & \ding{51} & Accountability & Long-form QA \\
WebGPT~\cite{webgpt} & GPT-3 & 175B & \ding{51} & Accountability & Long-form QA \\
\rowcolor{lightgray}WebBrain~\cite{webbrain} & BART & 400M & \ding{51} & Accountability & Long-form QA \\
RARR~\cite{RARR} & PaLM & 540B & $\times$ & Accountability & ODQA, Reasoning, Conversational QA \\
\rowcolor{lightgray}SearChain~\cite{Search-in-the-Chain} & GPT-3.5 & - & $\times$ & Accountability & Knowledge-intensive Tasks \\
\midrule
P2Bot~\cite{p2bot} & Transformer & - & \ding{51} & Personalization & Personalized Dialogue \\
\rowcolor{lightgray}P-Soups~\cite{PersonalizedSoups} & Tulu & 7B & \ding{51} & Personalization & Personalized Dialogue \\
OPPU~\cite{OPPU} & Llama-2 & 7B & \ding{51} & Personalization & Language Model Personalization Tasks \\
\rowcolor{lightgray}Zhongjing~\cite{Yang2024Zhongjing} & Ziya-LLaMA & 13B & \ding{51} & Healthcare & Chinese Medical Dialogue \\
Mental-LLM~\cite{xu2023Mental-LLM} & Alpaca/GPT-3.5/... & 7B/-/... & \ding{51}/$\times$ & Healthcare & Mental Health Reasoning Tasks \\
\rowcolor{lightgray}Edu-Chat~\cite{dan2023Educhat} & LLaMA & 13B & \ding{51} & Education & ODQA, Education Tasks \\
\bottomrule
\end{tabular}
\end{table*}

\subsubsection{Knowledge Updating}
\label{sec:knowledge_updating}

In real-life scenarios, information is constantly evolving, and therefore, the GenIR system needs to continuously acquire the latest knowledge to meet users' information needs. Since the model's knowledge storage is limited, knowledge updating is necessary to ensure more reliable generated responses. In this section, we will discuss existing methods for knowledge updating from two perspectives: incremental learning and knowledge editing.

\textbf{(1) Incremental Learning} Incremental learning refers to the ability of machine learning models to continuously learn new skills and tasks while retaining previously acquired knowledge~\cite{survey_continual_learning, wu2402survey_CL, survey_knowledge_edit_LLMs, zhang2401Study_KE}. In the GenIR system, it is crucial to enable the language model to memorize the latest information while preventing the forgetting of previous knowledge.

One approach is \textit{Incremental Pre-training}, which does not rely on supervised data but continues pre-training on continuously updated corpora to alleviate catastrophic forgetting. For example, Baidu proposed the ERNIE 2.0 framework~\cite{Sun2020Ernie2.0}, enhancing language understanding through continuous multi-task learning. Jang et al.~\cite{Jang2021CKL} introduced Continual Knowledge Learning (CKL) to explore how LLMs update and retain knowledge amidst rapidly changing information, creating benchmarks like FUAR. Cossu et al.~\cite{Cossu2022continual_pretraining} studied continual pre-training for language and vision, finding that self-supervised or unsupervised methods are more effective in retaining previous knowledge compared to supervised learning. Additionally, Ke et al.~\cite{ke2302DAP} proposed Domain Adaptive Pre-training (DAP-training) to improve the model's adaptability to new domains while preventing forgetting using techniques like soft masking and contrastive learning. For domain-specific model construction, Xie et al.~\cite{Xie2311FinPythia-6.9B} introduced FinPythia-6.9B, an efficient continual pre-training method specifically designed for large-scale language models in the financial domain.

On the other hand, \textit{Incremental Fine-tuning} utilizes only labeled data for training. Progressive Prompts~\cite{Razdaibiedina2301Progressive_prompts} appends new soft prompts for each new task, facilitating knowledge transfer and reducing forgetting. DynaInst~\cite{mok2023DynaInst} enhances lifelong learning in pre-trained language models through parameter regularization and experience replay, employing dynamic instance and task selection for efficient learning under resource constraints. Jang et al.~\cite{jang2023exploring} challenge traditional multi-task prompt fine-tuning by refining expert models on individual tasks. Suhr et al.~\cite{suhr2024continual} introduce a feedback-driven continual learning approach for instruction-following agents, where natural language feedback is converted into immediate rewards via contextual bandits to optimize learning. O-LoRA~\cite{wang2310O-LoRA} achieves superior continual learning by training new tasks in orthogonal low-rank subspaces, significantly minimizing task interference. Peng et al.~\cite{Peng2404JARe} propose a scalable language model that dynamically adjusts parameters based on task requirements, effectively preventing the forgetting of previously learned tasks.

\textbf{(2) Knowledge Editing} Knowledge editing refers to the process of modifying and updating existing knowledge within language models~\cite{survey_knowledge_edit_LLMs, survey_knowledge_edit_NN}, distinct from incremental learning that focuses on adapting to new domains or tasks. By editing the weights or layers of a model, knowledge editing methods can correct erroneous facts and incorporate new knowledge, making it important before deploying GenIR systems. There are primarily three paradigms for internal knowledge editing within language models: adding trainable parameters, locate-then-edit, and meta-learning.

One method of \textit{Adding Trainable Parameters} is by integrating new single neurons (patches) in the final feed-forward neural network (FFN) layer, as in T-Patcher~\cite{T-Patcher} and CaliNet~\cite{CaliNet}, which serve as trainable parameters to adjust the model’s behavior. Alternatively, discrete code-book modules are introduced in the middle layers of the language model, as in GRACE~\cite{Aging_with_GRACE}, to adjust and correct information.

Moreover, the \textit{Locate-then-Edit} method first identifies the parameters corresponding to specific knowledge and then updates these targeted parameters directly. Common techniques involve identifying key-value pairs in the FFN matrix, known as "knowledge neurons," and updating them~\cite{Dai2022Knowledge_Neurons}. Techniques like ROME~\cite{ROME} use causal mediation analysis to pinpoint areas needing editing, and MEMIT~\cite{MEMIT} builds on ROME to implement synchronized editing in various scenarios. Methods such as PMET~\cite{PMET} employ attention mechanisms for editing, while BIRD~\cite{BIRD} introduces a bidirectional inverse relation modeling approach.

\textit{Meta-Learning}, another paradigm, uses hyper-networks to generate the necessary updates for model editing. KE (Knowledge Editor)~\cite{KE} predicts weight updates for each data point using a hyper-network. MEND~\cite{MEND}, by taking low-order decomposition of gradients as input, learns to rapidly edit language models to enhance performance. Additionally, MALMEN~\cite{MALMEN} separates the computations of hyper-networks and language models, facilitating the editing of multiple facts under a limited memory budget. These meta-learning mechanisms enable models to swiftly adapt to new knowledge and tasks.
A detailed comparison of representative reliable response generation methods is provided in Table~\ref{tab:gen_works}.

\subsection{External Knowledge Augmentation}
\label{sec:external_knowledge_enhance}

Although large language models have demonstrated significant effectiveness in response generation, issues such as susceptibility to hallucinations, difficulty handling in-domain knowledge, and challenges with knowledge updating persist. Augmenting the model's generative process with external knowledge sources can serve as an effective way to tackle these issues. Based on the form of external knowledge employed, these approaches can be classified into retrieval augmentation and tool augmentation.
The blue area in Figure~\ref{fig:gen_modules} provides an overview of this section.

\subsubsection{Retrieval Augmentation}
\label{sec:retrieval_augmentation}

Retrieval-Augmented Generation (RAG) enhances the response of generative models by combining them with a retrieval mechanism~\cite{rag, fid, zhou2402MetaRAG}. By querying a large collection of documents, information that is relevant to the input query can be fetched and integrated into the input of the generative model. RAG enables generative models to be grounded in existing reliable knowledge, significantly improving the reliability of model generation. Typically, a RAG method involves a retriever and a generator. Based on the interaction flow between these two, RAG methods can be divided into four categories~\cite{ragsurvey}.

\textbf{(1) Sequential RAG:} Sequential RAG operates on a linear progression, where the retriever first retrieves relevant information and the generator utilizes this information to directly complete the response generation process.

The basic form of sequential RAG is a ``Retrieve-Read'' framework~\cite{rrr}, where early works perform joint~\cite{rag, realm, retro} or separate~\cite{fid} training of retriever and generator but require costly pre-training. In-Context RALM~\cite{in-context-rag} addresses this by directly using retrieved documents as input, leveraging the model's in-context learning without additional training.

With the widespread adoption of LLMs, most subsequent works are built on the foundation of a frozen generator. AAR~\cite{aar} fine-tunes a general retriever to adapt to the information acquisition preferences of the generative model. LLM-embedder~\cite{llm-embedder} uses rewards produced by LLM to train an embedding model dedicated to retrieval augmentation. ARL2~\cite{arl2} leverages LLM to annotate relevance scores in the training set and trains a retriever using contrastive learning.

Several works introduce pre-retrieval and post-retrieval processes~\cite{ragsurvey} into the sequential pipeline to enhance the overall efficiency. In the pre-retrieval process, the RRR model~\cite{rrr} introduces a rewriter module before the retriever, trained using the generator's feedback to enable the retrieval system to provide more suitable information for generation.

In the post-retrieval process, information compressors are proposed to filter out irrelevant content from documents, avoiding misleading the generator's response~\cite{lostinthemiddle, powerofnoise, jin2402BIDER}. RECOMP~\cite{recomp} uses both abstractive and extractive compressors to generate concise summaries of retrieved documents. LLMLingua~\cite{llmlingua} retains important tokens by calculating token importance based on the perplexity provided by the generative model. LongLLMLingua~\cite{longllmlingua} introduces query-aware compression and reranks retrieved documents based on importance scores to alleviate the ``loss in the middle'' phenomenon~\cite{lostinthemiddle}. PRCA~\cite{prca} employs reinforcement learning to train a text compressor adaptable to black-box LLMs and various retrievers, serving as a versatile plug-in.

\textbf{(2) Branching RAG:} In the Branching RAG framework, the input query is processed across multiple pipelines, and each pipeline may involve the entire process in the sequential pipeline. The outputs from all pipelines are merged to form the final response, allowing for finer-grained handling of the query or retrieval results.

In the pre-retrieval stage, TOC~\cite{toc} uses few-shot prompting to recursively decompose complex questions into clear sub-questions in a tree structure, retrieving relevant documents for each and generating a comprehensive answer. BlendFilter~\cite{blendfilter} enhances the original query using prompts with internal and external knowledge, retrieves related documents with the augmented queries, and merges them for a comprehensive response.

In the post-retrieval stage, REPLUG~\cite{replug} processes each retrieved document with the query through the generator separately and combines the resulting probability distributions to form the final prediction. GenRead~\cite{genread} prompts LLM to generate related documents and merges them with retrieved documents from the retriever as input, enhancing content coverage.

\textbf{(3) Conditional RAG:} The Conditional RAG framework adapts to various query types through distinct processes, improving the system's flexibility. Since there can be knowledge conflict between the knowledge from retrieved documents and the generator's own knowledge, RAG's effectiveness isn't consistent across all scenarios. To address this, common conditional RAG methods include a decision-making module that determines whether to engage the retrieval process for each query.

SKR~\cite{skr} trains a binary classifier on a dataset of questions LLMs can or cannot answer, determining at inference whether to use retrieval. Training labels are obtained by prompting the model to assess if external knowledge is needed. Self-DC~\cite{selfdc} uses the model's confidence score to decide on retrieval necessity, categorizing queries into unknown, uncertain, and known, with unknown queries processed through sequential RAG and uncertain ones decomposed into sub-questions. Rowen~\cite{rowen} introduces a multilingual detection module that perturbs the original question and measures response consistency to decide on retrieval.

\textbf{(4) Loop RAG:} Loop RAG involves deep interactions between the retriever and generator components. Owing to multi-turn retrieval and generation processes, accompanied by comprehensive interactions, it excels at handling complex and diverse input queries, yielding superior results in response generation.

ITER-RETGEN~\cite{iterretgen} introduces an iterative framework alternating between retrieval-augmented generation and generation-augmented retrieval, repeating this process to produce the final answer. IR-COT~\cite{ircot} follows a similar procedure to ITER-RETGEN but the iteration pauses based on the model's own generative process. FLARE~\cite{flare} conducts concurrent retrieval during generation, evaluating the need for retrieval at each new sentence based on the LLM's confidence score, dynamically supplementing information to enhance content reliability. COG~\cite{copyisallyouneed} models generation as continual retrieval and copying of segments from an external corpus, with the generator producing conjunctions to maintain fluency. Self-RAG~\cite{self-rag} adds special tokens into the vocabulary, allowing the generator to decide on retrieval, document importance, and whether to perform a critique.

Some works focus on deconstructing complex inquiries into sub-questions, addressing these individually to produce a more dependable response. \cite{self-ask} guides LLM to decompose complex questions into sub-questions, answer each using retrieved results, and synthesize the answers; RET-Robust~\cite{ret-robust} builds upon this by incorporating an NLI model to verify retrieved documents support the sub-question answers, reducing misinformation.

\subsubsection{Tool Augmentation}
\label{sec:tool_augmentation}

Although retrieval-augmented techniques have significantly improved upon the blind spots of a generator's self-knowledge, these methods struggle with the rapid and flexible update of information since they rely on the existence of information within an external corpus of documents. Tool augmentation, on the other hand, excels in addressing this issue by invoking various tools that allow for the timely acquisition and usage of the latest data, including finance, news, and more. Moreover, tool augmentation expands the scope of responses a model can offer, such as language translation, image generation, and other tasks, to more comprehensively meet users' information retrieval needs.

There are four categories of tools that can be utilized to construct a more reliable information retrieval system:

\textbf{(1) Search Engine:} Common search engine tools like Google Search and Bing Search help answer frequent and time-sensitive queries effectively. Self-Ask~\cite{self-ask} initially decomposes complex questions into multiple sub-questions, then uses a search engine to answer each sub-question, and finally generates a comprehensive answer to the complex question. ReAct~\cite{react} embeds search engine calls into the model's reasoning process, allowing the generative model to determine when to make calls and what queries to input for more flexible reasoning. New Bing can automatically search relevant information from Bing based on user input, yielding reliable and detailed answers, including citation annotations in the generated content.

Some works have also built advanced conversational systems based on tools like search engines. Internet-Augmented Generation~\cite{internet} enhances the quality of conversational replies by using search engines during conversations. LaMDA~\cite{lamda} and BlenderBot~\cite{blenderbot} combine search engines with conversational agents, constantly accessing internet information to enrich conversation factualness. WebGPT~\cite{webgpt} and WebCPM~\cite{webcpm} directly teach models to perform human-like browser operations by generating commands such as Search, Click, and Quote, facilitating the automated retrieval and acquisition of information.

\textbf{(2) Knowledge Graph (KG):} Compared to search engines, KGs are particularly useful for extracting structured, explicit knowledge. Relevant knowledge from a knowledge graph can be extracted and used as a prompt input to enhance the generative process~\cite{tog}. StructGPT~\cite{structgpt} introduces an iterative reading-and-reasoning framework where the model can access a knowledge graph through a well-designed interface, continually acquiring information and reasoning until an answer is obtained. RoG~\cite{rog} generates plausible reasoning paths from a KG, executes them in parallel, and integrates outcomes for a final answer; ToG~\cite{tog} allows the model to explore entities and links without pre-planning paths, continuously assessing reasoning feasibility.

\textbf{(3) API-based Tools:} An important part of the tools is the real-world APIs, which enable the model to obtain information from specific data sources, such as real-time stock information, movie services, code interpreters, and so on. However, the multitude and diversity of APIs, coupled with the adherence to certain operational protocols, make the teaching of API usage to models a focal point of this area.

Toolformer~\cite{toolformer} trains language models in a self-supervised manner to automatically call APIs when needed, using prompts to generate API calls, executing them, and filtering ineffective ones to form the dataset. Training with standard language modeling objectives yields models that can autonomously invoke APIs across tasks without losing language modeling capabilities. RestGPT~\cite{restgpt} formulates a framework prompting LLMs to invoke RESTful APIs, comprising an online planner, an API selector, and an executor. ToolLLM~\cite{ToolLLM} uses a large corpus of scraped APIs to build a dataset for fine-tuning. Gorilla~\cite{gorilla} introduces an information retriever providing the model with reference API documentation, facilitating retrieval-based information utilization during fine-tuning. ToolkenGPT~\cite{ToolkenGPT} incorporates each tool as a new token into the vocabulary, enabling the model to invoke APIs during inference as naturally as generating text.

Beyond learning to invoke APIs, CREATOR~\cite{creator} prompts models to write code based on actual problems as new tool implementations, with generated tools functioning through a code interpreter and demonstrating impressive results on complex tasks.

Some works additionally support multimodal inputs, broadening the application scope of the models. AssistGPT~\cite{assistgpt} offers a framework including modules like Planner, Executor, Inspector, and Learner, utilizing language and code for intricate inference. ViperGPT~\cite{vipergpt} feeds CodeX with user queries and visual API information to generate Python code invoking APIs, successfully completing complex visual tasks.

\textbf{(4) Model-based Tools:} With the swift expansion of diverse AI communities (i.e., Huggingface, ModelScope, GitHub), various types of AI models have become readily accessible for use, serving as a pivotal tool in enhancing generative retrieval systems. These AI models encompass a wide array of tasks, each accompanied by comprehensive model descriptions and usage examples.

HuggingGPT~\cite{hugginggpt} employs ChatGPT as a controller to deconstruct user queries into tasks, determining which models to invoke for execution. Similarly, Visual ChatGPT~\cite{visualgpt} integrates a visual foundation model with LLMs, leveraging ChatGPT as a prompt manager to mobilize visual foundation models like BLIP~\cite{blip} and ControlNet~\cite{controlnet}, adept at processing image-based requests efficiently compared to multi-modal models.

\subsection{Generating Response with Citation}
\label{sec:response_with_citation}
\begin{wrapfigure}{r}{0.5\textwidth}
\centering
\includegraphics[width=1\linewidth]{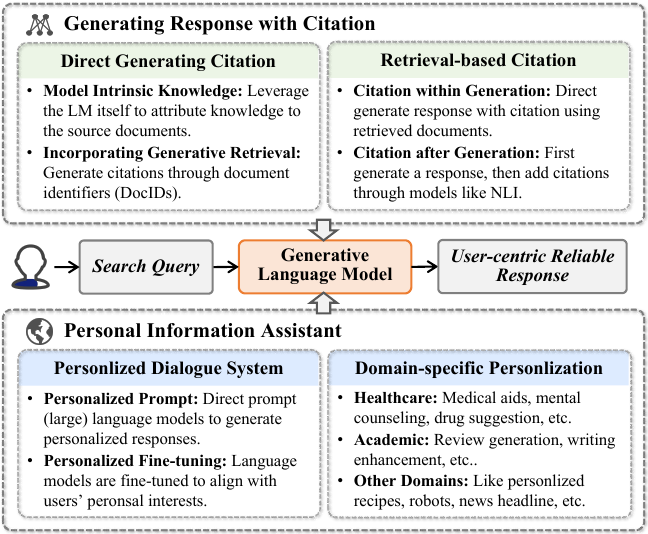}
\caption{
Generating response with citation and personal information assistant are also crucial approaches for building a reliable and user-centric GenIR system.
}
\label{fig:gen_modules_2}
\end{wrapfigure}

To build a reliable GenIR system, generating responses with citations is a promising approach~\cite{metzler2021rethinking, Huang2023Citation_A_Key, Establish_Trustworthiness}. Citations allow users to clearly understand the source of each piece of knowledge in the response, enhancing trust and facilitating widespread adoption. Existing methods can be divided into directly generating responses with citations and using a retrieval module to enhance the generated content.
Refer to the green portion in Figure~\ref{fig:gen_modules_2} for an overview of this section.

\subsubsection{Direct Generating Response with Citation}
\label{sec:direct_citation}
This method uses the model's intrinsic memory to generate source citations without relying on a retrieval module.

\textbf{(1) Model Intrinsic Knowledge}. 
Leveraging the capabilities of the language model itself, according-to prompting~\cite{According_to_Prompting} guides LLMs to more accurately cite information from pre-training data by adding phrases like "according to Wikipedia" in the prompts.

To improve citation quality, Iterative Feedback Learning (IFL)~\cite{IFL} employs a critique model to assess and provide feedback on generated text, iteratively enhancing LLMs' citation accuracy, content correctness, and fluency. Additionally, Fierro et al.~\cite{Fierro2024Plan_and_Citation} introduce a plan-based approach using a series of questions as a blueprint for content generation, with abstract and extractive attribution models, showing that planning improves citation quality.

\textbf{(2) Incorporating Generative Retrieval}.
As envisioned by Metzler et al.~\cite{metzler2021rethinking}, allowing the model to directly generate responses with citations is a promising approach for building an expert-level reliable IR system. Users receive reliable responses tailored to their needs without searching through returned documents. Moreover, the cited document is generated by the model through the generative retrieval approach described in Section~\ref{sec:generative_retrieval}, directly producing corresponding DocIDs.

Utilizing generative retrieval, 1-PAGER~\cite{1-PAGER} combines answer generation and evidence retrieval by generating N-gram DocIDs through constrained decoding using FM-Index~\cite{fm_index}, enabling step-by-step corpus partitioning, document selection, and response generation. This method matches retrieval-then-read methods in accuracy and surpasses closed-book QA models by attributing predictions to specific evidence, offering a new scheme for integrating retrieval into seq2seq generation.

Recently,~\cite{Khalifa2024SourceAwareTE} proposes a source-aware training method where models learn to associate DocIDs with knowledge during pre-training and provide supporting citations during instruction tuning, effectively achieving knowledge attribution and enhancing LLM verifiability.

\subsubsection{Retrieval-based Response with Citation}
\label{sec:retrieval_based_citation}

To enhance the accuracy of citations, several methods have been developed based on retrieval techniques to fetch relevant documents, thereby improving the quality of responses with embedded citations.

\textbf{(1) Citation within Generation}.
Following retrieval, models directly generate responses that include citations. Initially, systems like WebGPT~\cite{webgpt}, LaMDA~\cite{lamda}, and WebBrain~\cite{webbrain} utilized web pages or Wikipedia to construct large-scale pre-training datasets, teaching models how to generate responses with citations.

Subsequently, more advanced strategies for citation generation were proposed. For instance, Search-in-the-Chain (SearChain)~\cite{Search-in-the-Chain} first generates a reasoning chain (Chain-of-Query, CoQ) via LLM prompts, then interacts with each CoQ node using retrieval for verification and completion, ultimately generating the reasoning process and marking citations at each inference step.

LLatrieval~\cite{LLatrieval} suggests continuously improving retrieval results through iterative updating, verifying whether retrieved documents support the generated answers until satisfaction. AGREE~\cite{AGREE} uses a Natural Language Inference (NLI) model to verify consistency between LLM-generated answers and retrieved documents, employing a Test-Time Adaptation (TTA) strategy that allows LLMs to actively search and cite current information during generation, enhancing response accuracy and reliability. VTG~\cite{VTG} integrates an evolved memory system and a dual-layer validator for generating verifiable text, combining long-term and short-term memories to adapt to dynamic content, and uses an NLI model to evaluate logical support between claims and evidence.

Based on the Graph of Thoughts (GoT)~\cite{GoT}, HGOT~\cite{HGOT} improves context learning in retrieval-augmented settings by constructing a hierarchical GoT, leveraging the LLM's planning capabilities to break down complex queries into smaller sub-queries and introducing a scoring mechanism to assess the quality of retrieved paragraphs.

Employing reinforcement learning, Huang et al.~\cite{Huang2024Citation_via_Rewards} introduce a fine-grained reward mechanism to train language models, allocating specific rewards for each generated sentence and citation to teach models accurate external source citation. This approach uses rejection sampling and reinforcement learning algorithms to enhance citation-inclusive text generation through localized reward signals. APO~\cite{APO} reimagines attributive text generation as a preference learning problem, automatically generating preference data pairs to reduce annotation costs, and uses progressive preference optimization and experience replay to reinforce preference signals without overfitting or text degradation.

\textbf{(2) Citation after Generation}.
This approach involves models first generating a response, then adding citations through models like NLI. RARR~\cite{RARR} improves attributability by automatically finding external evidence for the language model's output and post-editing to correct content while preserving the original output, enhancing attribution capabilities without altering the existing model. PURR~\cite{PURR} employs an unsupervised learning method where LLMs generate text noise themselves, then trains an editor to eliminate this noise, improving attribution performance and significantly speeding up generation. CEG~\cite{CEG} searches for supporting documents related to generated content and uses an NLI-based citation generation module to ensure each statement is supported by citations. "Attribute First, then Generate"~\cite{Attribute_First_then_Generate} decomposes the generation process, first selecting relevant source text details and then generating based on these details, achieving localized attributability with each sentence supported by a clear source, greatly reducing manual fact-checking workload.

\subsection{Personal Information Assistant}
\label{sec:personal_information_assistant}
The core of the GenIR system is the user, so understanding user intent is crucial. Researchers have explored various methods like personalized search~\cite{zhou2020enc_his, wang2023subtopic, zhou2024cog_per_search, liu2023how_to_personalize}, dialogue~\cite{Zhang2018I_have_a_dog, Ma2021DHAP, p2bot}, and recommender~\cite{Liu2304ChatGPT_Rec, Dai2305ChatGPT_Rec, 2305_BookGPT} systems to explore users' interests. Specifically, personalized information assistants aim to better understand users' personalities and preferences, generating personalized responses to better meet their information needs. This section reviews the progress in research on personalized dialogue and domain-specific personalization.
An overview of this section is provided in the blue area of Figure~\ref{fig:gen_modules_2}.

\subsubsection{Personalized Dialogue System}
\label{sec:direct_personalized}
To better understand user needs, researchers have explored two main approaches: personalized prompt design and model fine-tuning.

\textbf{(1) Personalized Prompt}.
For personalized prompt design, Liu et al.~\cite{Liu2304ChatGPT_Rec} and Dai et al.~\cite{Dai2305ChatGPT_Rec} input users' interaction and rating history into ChatGPT~\cite{chatgpt} for in-context learning, effectively generating personalized responses. LaMP~\cite{LaMP} enhances the language model's personalized output by retrieving personalized history from user profiles. Using long-term history, \cite{2305_LLM_Journey} designs prompts describing users' long-term interests, needs, and goals for input into LLMs. BookGPT~\cite{2305_BookGPT} uses LLM prompts, interactive querying methods, and result verification frameworks to obtain personalized book recommendations. PerSE~\cite{2310_personalized_story_eval} infers preferences from several reviews by a specific reviewer and provides personalized evaluations for new story inputs.

Using prompt rewriting, \cite{2310_Prompt_Rewriting_for_Personalized} proposes a method combining supervised and reinforcement learning to better generate responses from frozen LLMs. Similarly, \cite{2310_Tailored_Visions} rewrites user input prompts using extensive user text-to-image interaction history to align better with expected visual outputs.

\textbf{(2) Personalized Fine-tuning}.
This line of work focuses on fine-tuning models for personalized response generation. Zhang et al.~\cite{Zhang2018I_have_a_dog} introduced the Persona-Chat dataset with 5 million personas to train models for more personalized dialogues. Mazar{\'{e}} et al.~\cite{Mazare2018_TrainingMillionsof} created a dataset of over 700 million conversations extracted from Reddit, demonstrating the effectiveness of training dialogue models on large-scale personal profiles. $\mathcal{P}^{2}$Bot~\cite{p2bot} generates personalized and consistent dialogues by simulating the perception of personalities between conversation participants. DHAP~\cite{Ma2021DHAP} designs a novel Transformer structure to automatically learn implicit user profiles from dialogue history without explicit personal information. Wu et al.~\cite{Wu2021Personalized_Response_Generation} propose a generative segmentation memory network to integrate diverse personal information. Fu et al.~\cite{Fu2022aThousandHamlets} developed a variational approach to model the relationship between personal memory and knowledge selection, with a bidirectional learning mechanism.

Using reinforcement learning, Cheng et al.~\cite{Cheng2309Everyone_Deserves_a_Reward} collected a domain-specific preference (DSP) dataset and proposed a three-stage reward model learning scheme, including base model training, general preference fine-tuning, and customized preference fine-tuning. Jang et al.~\cite{PersonalizedSoups} developed "Personalized Soups," first optimizing multiple policy models with different preferences using PPO~\cite{PPO}, then dynamically combining parameters during inference.

Using retrieval-enhanced methods, LAPDOG~\cite{Huang2023LAPDOG} retrieves relevant information from story documents to enhance personal profiles and generate better personalized responses. SAFARI~\cite{SAFARI} leverages LLMs' planning and knowledge integration to generate responses consistent with character settings. Inspired by writing education, Li et al.~\cite{Li2308Teach_LLMs_to_Personalize} proposed a multi-stage, multi-task framework including retrieval, ranking, summarization, synthesis, and generation to teach LLMs personalized responses. For subjective tasks, \cite{Wozniak2402Personalized_LLMs} studied the superior performance of personalized fine-tuning in subjective text perception tasks compared to non-personalized models.

To achieve a personalized information assistant for every user, OPPU~\cite{OPPU} uses personalized PEFT~\cite{peft} to store user-specific behavioral patterns and preferences, showing superior performance. For multimodal scenarios, PMG~\cite{shen2404PMG} proposes a personalized multi-modal generation method that transforms user behavior into natural language, allowing LLMs to understand and extract user preferences.

\subsubsection{Domain-specific Personalization}
\label{sec:rag_personalized}
Understanding users' personalized information needs, the GenIR system has broad applications across various domains such as healthcare, academia, education, and recipes.

\textbf{(1) Healthcare}.
In AI-assisted healthcare, personalization plays a crucial role. Liu et al.~\cite{Liu2305LLM_zeroshot_health} utilize few-shot tuning to process time-series physiological and behavioral data. Zhang et al.~\cite{Zhang2307LLMasClinicalAssistant} implement medical diagnosis identification and diagnostic assistance using prompts from ChatGPT~\cite{chatgpt} and GPT-4~\cite{gpt4}. Yang et al.~\cite{Yang2024Zhongjing} propose an LLM for traditional Chinese medicine called Zhongjing, based on LLaMA~\cite{llama}, undergoing pre-training, supervised fine-tuning, and RLHF~\cite{rlhf}. Abbasian et al.~\cite{Abbasian2310openCHA} introduce an open-source LLM-based conversational health agent framework called openCHA, which collects necessary information through specific actions and generates personalized responses. MedAgents~\cite{Tang2311MedAgents} propose a multidisciplinary collaboration framework where LLM-based agents engage in multi-round cooperative discussions to enhance expertise and reasoning.

For mental healthcare, Mental-LLM~\cite{xu2023Mental-LLM} presents a framework using LLMs to predict mental health from social media text data, with prompting-based and fine-tuning methods for real-time monitoring of issues like depression and anxiety. Lai et al.~\cite{Lai2307Psy-LLM} introduce Psy-LLM, a psychological consultation aid combining pre-trained LLMs with real psychologist Q\&As and psychological articles.

For medication suggestions, Liu et al.~\cite{liu2023Pharmacygpt} propose PharmacyGPT, a framework for generating personalized patient groups, formulating medication plans, and predicting outcomes.

\textbf{(2) Academic}.
In the academic domain, RevGAN~\cite{RevGAN} can automatically generate controllable and personalized user reviews based on users' emotional tendencies and stylistic information. For writing assistants, Porsdam et al.~\cite{porsdam2023AUTOGEN} explore personalized enhancement of academic writing using LLMs like GPT-3~\cite{gpt3}, showing higher quality after training with authors' published works. Similarly, to address the lack of personalized outputs in LLMs, Mysore et al.~\cite{Mysore2311Pearl} propose Pearl, a personalized LLM writing assistant trained on users' historical documents, and develop a KL divergence training objective for retrievers.

\textbf{(3) Education}.
Cui et al.~\cite{Cui2023Personalized_exercise} propose an adaptive and personalized exercise generation method that adjusts difficulty to match students' progress by combining knowledge tracing and controlled text generation. EduChat~\cite{dan2023Educhat} learns education-specific functionalities through pre-training on educational corpora and fine-tuning on customized instructions, addressing delayed knowledge updates and lack of expertise in LLMs.

\textbf{(4) Other Domains}.
For recipe generation tasks, Majumder et al.~\cite{Majumder2019PersonalRecipes} propose a personalized generation model based on users' historical recipe consumption, enhancing personalization. For personalized headline generation, Zhang et al.~\cite{Zhang2022Personalized_Headline_Generation} simulate users' interests based on browsing history to generate news headlines. Salemi et al.~\cite{LaMP} propose the LaMP benchmark, including personalized generation tasks like news headline, academic title, email subject, and tweet rewriting. Additionally, for personalized assistance with home cleaning robots, TidyBot~\cite{Wu2023TidyBot} uses LLMs to generalize from user examples to infer user preference rules.

\section{Evaluation}
\label{sec:evaluation}

This section will provide a range of evaluation metrics and benchmarks for generative information retrieval methods, along with analysis and discussions on their performance.

\subsection{Evaluation for Generative Document Retrieval}
\label{sec:eval_gr}

\subsubsection{Metrics}
\label{sec:gr_metrics}

In this section, we discuss several core metrics for evaluating Generative Retrieval (GR) methods. These metrics provide different perspectives on the effectiveness of a GR system, including its accuracy, efficiency, and the relevance of its results. Specifically, we consider Recall, R-Precision, Mean Reciprocal Rank (MRR), Mean Average Precision (MAP), and Normalized Discounted Cumulative Gain (nDCG). Each metric captures unique aspects of retrieval performance, allowing for a comprehensive assessment of the system's capabilities.

\begin{itemize}[leftmargin=*]
\item \textbf{Recall} measures the proportion of relevant documents retrieved by the search system, reflecting its ability to find all relevant items.
\item \textbf{R-Precision} evaluates the precision at a rank position corresponding to the number of relevant documents, balancing precision and recall at a specific cutoff.
\item \textbf{Mean Reciprocal Rank (MRR)} captures the average rank position of the first relevant document, emphasizing the system's ability to return relevant results early in the ranking.
\item \textbf{Mean Average Precision (MAP)} calculates the average precision across multiple queries, considering the exact positions of all relevant documents and providing a comprehensive measure of retrieval accuracy.
\item \textbf{Normalized Discounted Cumulative Gain (nDCG)} takes into account not only the relevance of the documents returned but also their positions in the result list, reflecting both the quality and the ordering of the results.
\end{itemize}

For detailed mathematical formulations of these metrics, please refer to Appendix~\ref{app:gr_metrics}.

\subsubsection{Benchmarks}
\label{sec:gr_benchmarks}

Evaluating the effectiveness of GR methods relies on high-quality and challenging benchmark datasets.

\textbf{MS MARCO}~\cite{ms_marco} is a large-scale dataset designed for machine reading comprehension, retrieval, and question-answering tasks in web search environments. It contains millions of documents and passages derived from real user queries, providing a realistic benchmark for assessing GR systems in complex search scenarios.

\textbf{Natural Questions (NQ)}~\cite{nq} is a question-answering dataset introduced by Google, utilizing Wikipedia as its primary corpus. It includes a vast number of natural user queries and their corresponding answers, making it suitable for evaluating the retrieval performance of GR systems in addressing real-world informational needs.

\textbf{KILT (Knowledge-Intensive Language Tasks)}~\cite{kilt} is a comprehensive benchmark integrating multiple categories of knowledge-intensive tasks such as fact checking, entity linking, slot filling, open-domain QA, and dialogue. Utilizing Wikipedia as its corpus, KILT aims to evaluate the effectiveness of information retrieval systems in handling complex language tasks that require extensive background knowledge.

\textbf{TREC Deep Learning Track 2019 \& 2020}~\cite{trec_19_dl, trec_20_dl} focus on leveraging deep learning techniques to enhance information retrieval efficiency, primarily through document and passage ranking tasks. These tracks utilize the MS MARCO dataset to emulate real-world search queries, providing a standardized environment for benchmarking various retrieval methodologies.

% It is noteworthy that, due to the challenge of indexing large-scale document corpus by GR models, some methods like~\cite{dsi, dynamicretriever, nci, dsiqg, ultron, chen_understand_dsi, lmindexer} have adopted the approach of evaluating retrieval performance on smaller subsets of documents, ranging from 10K to 300K, along with their corresponding training and test pairs.

\textbf{DynamicIR.}
For dynamic corpora, DynamicIR~\cite{dynamicir} proposes a task framework based on StreamingQA~\cite{StreamingQA} benchmark for evaluating IR models within dynamically updated corpora. Through experimental analysis, DynamicIR revealed that GR systems are superior in adapting to evolving knowledge, handling temporally informed data, and are more efficient in terms of memory, indexing time, and FLOPs compared to dense retrieval systems. 

\textbf{ExcluIR.}
For exclusionary retrieval tasks, where users explicitly indicate in their queries that they do not want certain information, ExcluIR~\cite{ExcluIR} provides a set of resources. This includes an evaluation benchmark and a training set to help retrieval models understand and process exclusionary queries.

For detailed descriptions and comprehensive information about benchmark datasets, please refer to Appendix~\ref{app:gr_benchmarks}.

\subsubsection{Analysis}
\label{sec:gr_analysis}

In addition to the benchmarks and metrics for evaluating the performance of GR methods, there is a series of works that have conducted detailed analyses and discussions to study the behavior of GR models.

\textbf{Understanding Generative Retrieval.}
To understand the performance of DSI~\cite{dsi} in text retrieval, Chen et al.~\cite{chen_understand_dsi} examines uniqueness, completeness, and relevance ordering. These respectively reflect the system's ability to distinguish between different documents, retrieve all relevant documents, and accurately rank documents by relevance. Experimental analysis find that DSI excels in remembering the mapping from pseudo queries to DocIDs, indicating a strong capability to recall specific DocIDs from particular queries. However, the study also pointed out DSI's deficiency in distinguishing relevant documents from random ones, negatively impacting its retrieval effectiveness.

Exploring the connection between generative and dense retrieval, \cite{gen_as_dense} demonstrates that they can be considered as bi-encoders in dense retrieval. Specifically, the authors analyze the computation of dot products during the generative retrieval process, which is similar to the calculation of dot products between query vectors and document vectors in dense retrieval. Following this, \cite{wu2024gr_as_mvdr} revisits generative retrieval from the perspective of multi-vector dense retrieval (MVDR), revealing a common framework in computing document-query relevance between the two methods. This work also analyzes their differences in document encoding and alignment strategies, further confirming through experiments the phenomenon of term matching in the alignment matrices and their commonalities in retrieval.

\textbf{Large-scale Experimental Analysis.}
Later, Pradeep et al.~\cite{pradeep_how_does} conduct the first comprehensive experimental study on GR techniques over large document sets, such as the 8.8M MS MARCO passages. It was found that among all the techniques examined, using generated pseudo queries to augment training data remains the only effective method on large document corpus. The strongest result in the experiments was achieved by using a training task that only utilized synthetic queries to Naive DocIDs, expanding the model to T5-XL (3B parameters) to achieve an MRR@10 of 26.7. Surprisingly, increasing the parameters to T5 XXL (11B) in the same setup did not improve performance but rather led to a decline. These findings suggest that more research and in-depth analysis are needed in the GR field, and possibly additional improvements to the paradigm, to fully leverage the power of larger language models.

\textbf{Out-of-distribution Perspective.}
For out-of-distribution (OOD) robustness of GR models, Liu et al.~\cite{gr-ood} investigate three aspects: query variations, new query types, and new tasks. Their study showed that all types of retrieval models suffer from performance drops with query variations, indicating sensitivity to query quality and structure. However, when dealing with new query types and tasks, GR models showed different levels of adaptability, with pre-training enhancing their flexibility. The research highlights the critical need for OOD robustness in GR models for dealing with ever-changing real-world information sources.

\subsubsection{Experiments}
\label{sec:gr_exp}

Analyzing experimental results is essential for understanding the performance of different GR models. This section provides a comprehensive evaluation of current GR models on widely used benchmark tests and examines their applicability and limitations in scenarios such as web search, question answering, and knowledge-intensive tasks. The overall results are presented in Table~\ref{tab:msnq_results} and Table~\ref{tab:kilt_results}.

\textbf{Experimental Settings.} Our evaluation is based on the MS MARCO~\cite{ms_marco}, NQ~\cite{nq}, and KILT~\cite{kilt} benchmarks, which are commonly used datasets for existing GR methods. For the MS MARCO dataset, following previous works~\cite{ultron, genret, autotsg}, we use the MS MARCO 300K subset, which contains 320k documents, 360k training instances, and 772 testing instances. For the NQ dataset, following~\cite{dsi, nci, genret, autotsg, glen}, we use the NQ320K subset, which, after deduplication based on titles, contains 109k documents, 320k training instances, and 7,830 testing instances. For the KILT benchmark, we use the standard development sets. Detailed statistics are available in previous works~\cite{kilt, corpusbrain}.

Regarding evaluation metrics, we employ Recall@\{1, 10, 100\} and MRR@\{10, 100\} for the MS MARCO and NQ datasets, and R-Precision for the KILT benchmark. In our comparisons, we include not only existing representative GR methods but also sparse retrieval methods such as BM25~\cite{bm25} and SPLADEv2~\cite{spladev2}, which are based on bag-of-words representations, and dense retrieval methods like DPR~\cite{dpr}, GTR~\cite{gtr}, RAG~\cite{rag} and MT-DPR~\cite{mtdpr}, which rely on dense embeddings.

Due to variations in datasets, corpus sizes, and evaluation metrics across different methods, alignment is necessary for a fair comparison. For the methods evaluated in our experiments, we primarily use results reported in existing papers. For methods where settings are not aligned, we provide results based on our own implementations. 

\begin{table*}[!t]
\centering
\caption{Overall retrieval performance on the MS MARCO (300K) and Natural Questions (320K) Datasets. The best results are \textbf{Bold} and the second-best are \underline{Underlined}. Symbol "*" indicates results from our own implementation, while other results are consistent with those reported in existing papers.} 
\vspace{-2mm}
\setlength\tabcolsep{3.2pt}
\fontsize{8pt}{10.1pt}\selectfont
\label{tab:msnq_results}
\begin{tabular}{lccccccccccc}
\toprule
\multirow{2.5}{*}{\textbf{Model}} & \multirow{2.5}{*}{\textbf{Doc Rep.}} & \multicolumn{5}{c}{\textbf{MS MARCO}} & \multicolumn{5}{c}{\textbf{Natural Questions (NQ)}} \\
\cmidrule(lr){3-7}\cmidrule(l){8-12}
& & R@1 & R@10 & R@100 & M@10 & M@100 & R@1 & R@10 & R@100 & M@10 & M@100 \\
\midrule
\multicolumn{11}{l}{\textit{Sparse\&Dense Retrieval}} \\
BM25~\cite{genret} & Bag-of-words & 0.196 & 0.591 & 0.861 & 0.313 & 0.325 & 0.297 & 0.603 & 0.821 & - & 0.402 \\
SPLADEv2~\cite{autotsg} & Bag-of-words & 0.328 & 0.779 & \underline{0.956} & 0.443 & 0.452 & 0.624 & 0.873 & \underline{0.954} & 0.726 & 0.731 \\
DPR~\cite{genret} & Dense Vector & 0.271 & 0.764 & 0.948 & 0.424 & 0.433 & 0.502 & 0.777 & 0.909 & - & 0.489 \\
GTR-Base~\cite{genret} & Dense Vector & 0.332 & \textbf{0.793} & \textbf{0.960} & 0.484 & 0.485 & 0.560 & 0.844 & 0.937 & - & 0.662 \\
\midrule
\multicolumn{11}{l}{\textit{Generative Retrieval}} \\
GENRE~\cite{autotsg} & Title & 0.266 & 0.579 & 0.751 & 0.361 & 0.368 & 0.591 & 0.756 & 0.814 & 0.653 & 0.656 \\
DSI~\cite{autotsg} & Semantic ID & 0.257 & 0.538 & 0.692 & 0.339 & 0.346 & 0.533 & 0.715 & 0.816 & 0.594 & 0.598 \\
DSI-QG~\cite{ultron,genret} & Semantic ID & 0.288 & 0.623 & - & 0.385 & - & 0.631 & 0.807 & 0.880 & - & 0.695 \\
NCI~\cite{genret} & Semantic ID & 0.301 & 0.643 & 0.851 & 0.408 & - & 0.659 & 0.852 & 0.924 & - & 0.731 \\
SEAL~\cite{genret} & Sub-string & 0.259 & 0.686 & 0.879 & 0.393 & 0.402 & 0.570 & 0.800 & 0.914 & - & 0.655 \\
Ultron~\cite{autotsg} & Title+URL & 0.304 & 0.676 & 0.794 & 0.432 & 0.437 & 0.654 & 0.854 & 0.911 & 0.726 & 0.729 \\
GenRet~\cite{genret} & Learnable & - & - & - & - & - & 0.681 & 0.888 & 0.952 & - & 0.759 \\
MINDER~\cite{autotsg} & Multi-view & 0.289 & 0.728 & 0.916 & 0.431 & 0.435 & 0.627 & 0.869 & 0.933 & 0.709 & 0.713 \\
LTRGR* & Multi-view & 0.327 & 0.759 & 0.929 & 0.463 & 0.469 & 0.644 & 0.879 & 0.941 & 0.721 & 0.726 \\
GLEN~\cite{glen} & Learnable & - & - & - & - & - & 0.691 & 0.860 & - & - & 0.754 \\
TSGen~\cite{autotsg} & Term Set & \textbf{0.384} & \underline{0.781} & 0.931 & \textbf{0.502} & \textbf{0.505} & \textbf{0.708} & \underline{0.889} & 0.948 & \textbf{0.771} & \textbf{0.774} \\
NOVO~\cite{novo} & Term Set & - & - & - & - & - & \underline{0.693} & \textbf{0.897} & \textbf{0.959} & - & \underline{0.767} \\
DGR* & Multi-view & \underline{0.359} & 0.779 & 0.934 & \underline{0.498} & \underline{0.504} & 0.682 & 0.887 & 0.949 & \underline{0.759} & 0.764 \\
\bottomrule
\end{tabular}
\end{table*}

\textbf{Results on MS MARCO and NQ Datasets.}
MS MARCO and Natural Questions (NQ) are among the most widely used benchmarks for evaluating generative retrieval (GR) methods, particularly in the contexts of web search and question answering. Table~\ref{tab:msnq_results} presents a detailed comparison of various GR models against traditional sparse and dense retrieval methods on these datasets.

(1) Overall Performance Comparison. Overall, GR methods demonstrate competitive performance compared to sparse and dense retrieval baselines. Specifically, on the MS MARCO dataset, GR models such as {TSGen} and {DGR} achieve Recall@1 scores of 0.384 and 0.359 respectively, surpassing dense methods like DPR (0.271) and being comparable to SPLADEv2 (0.328). On the NQ dataset, GR models also show strong performance, with {TSGen} attaining the highest Recall@1 of 0.708, outperforming both SPLADEv2 (0.624) and DPR (0.502).

(2) Term Set DocID Methods. Analyzing models that utilize term set-based document identifiers, such as {TSGen} and {NOVO}, reveals that these methods excel in both datasets. {TSGen} leads with the highest Recall@1 and MRR@10 on MS MARCO and NQ respectively, indicating robust retrieval capabilities. {NOVO} also performs exceptionally well on the NQ dataset, achieving the second-best Recall@1 and MRR@10, demonstrating the effectiveness of term set representations in capturing relevant document information.

(3) Multi-view DocID Methods. Multi-view approaches, exemplified by {MINDER}, {LTRGR}, and {DGR}, show consistent improvements over several metrics. For instance, {LTRGR} achieves the highest Recall@10 on MS MARCO (0.759) and maintains strong performance across other metrics and on the NQ dataset. These results suggest that leveraging multi-view DocIDs, ranking and distillation training methods enhances retrieval effectiveness by capturing diverse aspects of the documents.

(4) Learnable DocID Methods. Learnable DocID models, such as {GenRet} and {GLEN}, exhibit mixed performance. While {GenRet} shows competitive Recall@1 on NQ (0.681), it does not report results on MS MARCO. {GLEN} achieves the highest MRR@100 on NQ (0.754) but lags behind in other metrics. This indicates that learnable DocID approaches may benefit from further refinement to consistently outperform other representation methods across different datasets.

(5) Other DocID Methods. Other methods like {GENRE}, {DSI}, {NCI}, {SEAL}, and {Ultron}, generally underperform compared to term set and multi-view DocID methods. For example, on the MS MARCO dataset, {GENRE} achieves a Recall@1 of 0.266 and an MRR@10 of 0.361, which are significantly lower than {TSGen} (Recall@1 = 0.384, MRR@10 = 0.502) and {LTRGR} (Recall@1 = 0.327, MRR@10 = 0.463). 
The lower performance of methods utilizing simpler DocID designs (e.g. titles, semantic IDs) highlights the need for more sophisticated or alternative DocID strategies to effectively capture key information for high-quality retrieval across different scenarios.

\begin{table*}[!t]
\centering
\caption{Overall retrieval performance on the KILT Benchmark. The best results are \textbf{Bold} and the second-best are \underline{Underlined}. Symbol "*" indicates results from our own implementation, while other results are consistent with those reported in existing papers.}
\vspace{-2mm}
\setlength\tabcolsep{2.6pt}
\fontsize{8pt}{10.2pt}\selectfont
\label{tab:kilt_results}
\begin{tabular}{lcccccccccccc}
\toprule
\multirow{2.5}{*}{\textbf{Model}} & \multirow{2.5}{*}{\textbf{Doc Rep.}} & \textbf{FC} & \multicolumn{3}{c}{\textbf{Entity Linking}} & \multicolumn{2}{c}{\textbf{Slot Filling}} & \multicolumn{4}{c}{\textbf{Open Domain QA}} & \textbf{Dial.} \\
\cmidrule(lr){3-3}\cmidrule(lr){4-6}\cmidrule(lr){7-8}\cmidrule(lr){9-12}\cmidrule(lr){13-13}
& & \textbf{FEVER} & \textbf{AY2} & \textbf{WnWi} & \textbf{WnCw} & \textbf{TREx} & \textbf{zsRE} & \textbf{NQ} & \textbf{HoPo} & \textbf{TQA} & \textbf{ELI5} & \textbf{WoW} \\
\midrule
\multicolumn{13}{l}{\textit{Sparse\&Dense Retrieval}} \\
BM25~\cite{mtdpr} & Bag-of-words & 0.501 & 0.035 & - & - & 0.586 & 0.664 & 0.258 & 0.440 & 0.294 & - & 0.275 \\
RAG~\cite{kilt} & Dense Vector & 0.635 & 0.774 & 0.490 & 0.467 & 0.293 & 0.654 & 0.603 & 308 & 0.493 & 0.104 & 0.467 \\
MT-DPR~\cite{mtdpr} & Dense Vector & 0.747 & 0.838 & - & - & 0.692 & 0.772 & 0.615 & 0.442 & 0.620 & - & 0.397 \\
\midrule
\multicolumn{13}{l}{\textit{Generative Retrieval}} \\
BART* & Semantic ID & 0.003 & 0.001 & 0.000 & 0.000 & 0.000 & 0.001 & 0.000 & 0.000 & 0.000 & 0.000 & 0.000 \\
BART~\cite{corpusbrain} & Title & 0.819 & 0.892 & 0.676 & 0.623 & 0.752 & 0.911 & 0.586 & 0.487 & 0.676 & 0.121 & 0.510 \\
T5~\cite{kilt} & Title & - & 0.866 & 0.474 & 0.465 & - & - & - & - & - & - & - \\
GENRE~\cite{genre} & Title & \textbf{0.847} & \textbf{0.928} & \textbf{0.877} & \textbf{0.706} & \textbf{0.797} & \underline{0.948} & \underline{0.643} & \underline{0.518} & \underline{0.711} & \textbf{0.135} & \textbf{0.563} \\
SEAL* & Sub-string & \underline{0.826} & 0.866 & \underline{0.809} & 0.651 & 0.704 & 0.919 & \textbf{0.658} & \textbf{0.565} & \textbf{0.715} & 0.124 & 0.527 \\
CorpusBrain~\cite{corpusbrain} & Title & 0.821 & \underline{0.908} & 0.723 & \underline{0.662} & \underline{0.776} & \textbf{0.983} & 0.591 & 0.501 & 0.688 & \underline{0.129} & \underline{0.538} \\
\bottomrule
\end{tabular}
\end{table*}

\textbf{Results on KILT Benchmark.}
The KILT benchmark provides a comprehensive evaluation across various knowledge-intensive tasks, utilizing a large-scale Wikipedia corpus comprising 5.9 million documents. Overall results are shown in Table~\ref{tab:kilt_results}.

(1) Overall Performance Comparison. GR methods generally outperform traditional sparse and dense retrieval approaches in most tasks. Notably, {GENRE} achieves the highest scores in several categories, including FEVER (0.847), AY2 (0.928), WnWi (0.877), and WnCw (0.706), outperforming the best sparse method {BM25} and dense methods like {MT-DPR}.

(2) Title DocID Methods. Models utilizing title-based document identifiers consistently perform well on the KILT benchmark. For instance, {GENRE} and {BART} achieve FEVER scores of 0.847 and 0.821, respectively. This superior performance can be attributed to the fact that Wikipedia document titles accurately represent the key entities within each document, making the task of predicting titles relatively straightforward. Moreover, these models effectively leverage the pre-trained knowledge embedded within language models, enhancing their ability to generalize and retrieve relevant documents based on titles.

(3) Sub-string DocID Methods. Methods based on sub-string document identifiers also demonstrate strong performance on the KILT benchmark, particularly in question answering (QA) tasks. {SEAL} achieves the highest QA scores across several categories, including NQ (0.658), HoPo (0.565), TQA (0.715), and WoW (0.527). The ability of sub-string DocID methods to capture meaningful fragments of the documents likely contributes to their high accuracy in retrieving precise information necessary for answering questions effectively.

(4) DSI-based Numeric DocID Methods. In contrast, methods employing numeric Semantic DocIDs based on hierarchical k-means clustering~\cite{dsi} exhibit significantly diminished performance on the KILT benchmark. The {BART} model, which uses Semantic IDs and trained with just labeled queries, records scores close to zero across all tasks (e.g., FEVER: 0.003, AY2: 0.001). This decline is primarily due to the substantial increase in corpus size, and <query, document> pairs in training data cover only a small fraction of the entire document set. Consequently, these models struggle to generalize beyond the training pairs, just "memorizing" DocIDs without capturing the broader diversity of the corpus. This observation aligns with findings from \cite{pradeep_how_does}, which reported similar challenges of DSI~\cite{dsi} when scaling to an 8.8 million passage corpus in the MS MARCO benchmark.

\subsection{Evaluation for Response Generation}
\label{sec:eval_response}

\subsubsection{Metrics}
\label{sec:response_metrics}
Evaluating the quality of generated responses includes aspects such as accuracy, fluency, relevance, etc. In this section, we'll introduce the main metrics for evaluating reliable response generation, categorized into rule-based, model-based, and human evaluation metrics.

\textbf{(1) Rule-based Metrics}.
Exact Match (EM) is a straightforward evaluation method requiring the model's output to be completely identical to the reference answer at the word level. This full character-level matching is stringent, often used in tasks requiring precise and concise answers, such as question answering systems, e.g., NQ~\cite{nq}, TriviaQA~\cite{triviaqa}, SQuAD~\cite{squad}, etc. It simply calculates the ratio of perfectly matched instances to the total number of instances.

For the generation of longer text sequences, BLEU~\cite{bleu} is a common metric initially used to evaluate the quality of machine translation. It compares the similarity between the model's output and a set of reference texts by calculating the overlap of n-grams, thereby deriving a score. This method assumes that high-quality generation should have a high lexical overlap with the labeled answer.
Optimized from BLEU, METEOR~\cite{METEOR} is an alignment-based metric that considers not only exact word matches but also synonyms and stem matches. Additionally, METEOR introduces considerations for word order and syntactic structure to better assess the fluency and consistency of the generated text.

ROUGE~\cite{rouge}, also a commonly used metric for evaluating longer texts, by measuring the extent of overlap in words, sentences, n-grams, and so forth, between the generated text and a collection of reference texts. It focuses on recall, meaning it evaluates how much of the information in the reference text is covered by the generated text. ROUGE comes in various forms, including ROUGE-N, which evaluates based on n-gram overlap, and ROUGE-L, which considers the longest common subsequence, catering to diverse evaluation requirements.

Perplexity (PPL) is a metric for evaluating the performance of language models, defined as the exponentiation of the average negative log-likelihood, reflecting the model's average predictive ability for a given corpus of text sequences. The lower the perplexity, the stronger the model's predictive ability. Specifically, given a sequence of words $W = w_1, w_2, \ldots, w_N$, where $N$ is the total number of words in the sequence, PPL can be expressed as:
\begin{equation}
    \text{PPL}(W) = \exp\left\{-\frac{1}{N} \sum_{i=1}^{N} \log p(w_i | w_{<i})\right\},
\end{equation}
where $p(w_i | w_{<i})$ represents the pre-trained language model's probability of predicting the $i$-th word $w_i$ given the previous words $w_{<i}$.

\textbf{(2) Model-based Metrics}.
With the rise of pre-trained language models, a series of model-based evaluation metrics have emerged. These metrics utilize neural models to capture the deep semantic relationships between texts.

Unlike traditional rule-based metrics, BERTScore~\cite{bertscore} utilizes the contextual embeddings of BERT~\cite{bert} to capture the deep semantics of words, evaluating the similarity between candidate and reference sentences through the cosine similarity of embeddings. BERTScore employs a greedy matching strategy to optimize word-level matching and uses optional inverse document frequency weighting to emphasize important words, ultimately providing a comprehensive evaluation through a combination of recall, precision, and F1 score. BERTScore captures not only surface lexical overlap but also a deeper understanding of the semantic content of sentences.

Similarly based on BERT~\cite{bert}, BLEURT~\cite{BLEURT} designed multiple pre-training tasks, enhancing the model's ability to recognize textual differences with millions of synthetic training pairs. These pre-training tasks include automatic evaluation metrics (such as BLEU~\cite{bleu}, ROUGE~\cite{rouge}, and BERTScore~\cite{bertscore}), back-translation likelihood, textual entailment, etc. Each task provides different signals to help the model learn how to evaluate the quality of text generation.

BARTScore~\cite{bartscore}, based on the pre-trained seq2seq generative model BART~\cite{bart}, treats the evaluation of generated text as a text generation problem. Specifically, BARTScore determines the quality of text based on the transition probability between the generated text and reference text. BARTScore does not require additional parameters or labeled data and can flexibly evaluate generated text from multiple perspectives (such as informativeness, fluency, factuality, etc.) and further enhance evaluation performance through text prompts or fine-tuning for specific tasks.

FActScore~\cite{FActScore} focuses on the factual accuracy of each independent information point in long texts. It calculates a score representing factual accuracy by decomposing the text into atomic facts and verifying whether these facts are supported by reliable knowledge sources. This method provides a more detailed evaluation than traditional binary judgments and can be implemented efficiently and accurately through human evaluation and automated models (combining retrieval and powerful language models).

GPTScore~\cite{gptscore} is a flexible, multi-faceted evaluation tool that allows users to evaluate text using natural language instructions without the need for complex training processes or costly annotations. GPTScore constructs an evaluation protocol dynamically through task specification and aspect definition and utilizes the zero-shot capability of pre-trained language models to evaluate text quality, optionally using demonstration samples to improve evaluation accuracy.

\textbf{(3) Human Evaluation Metrics}.
Human evaluation is an important method for assessing the performance of language models, especially in complex tasks where automated evaluation tools struggle to provide accurate assessments. Compared to rule-based and model-based metrics, human evaluation is more accurate and reliable in real-world applications. This evaluation method requires human evaluators (such as experts, researchers, or everyday users) to provide comprehensive assessments of the model-generated content based on their intuition and knowledge.

Human evaluation measures the quality of language model outputs by integrating multiple assessment criteria, following~\cite{survey-eval}: Accuracy~\cite{human-accuracy} primarily evaluates the correctness of information and its correspondence with facts; Relevance~\cite{human-relevance} focuses on whether the model's output is pertinent to the specific context and user query; Fluency~\cite{human-fluency} examines whether the text is coherent, natural, and facilitates smooth communication with users; Safety~\cite{human-safety} scrutinizes whether the content may lead to potential adverse consequences or harm. These indicators collectively provide a comprehensive assessment of the model's performance in real-world settings, ensuring its effectiveness and applicability.

However, human evaluation also faces numerous challenges, primarily including high costs and time consumption, difficulty in controlling evaluation quality, inconsistency in evaluation dimensions, issues of consistency due to evaluators' subjectivity, and the need for professional evaluators for specific tasks. These problems limit the widespread application of human evaluation and the comparability of results~\cite{2006_eval_nlg}.

\subsubsection{Benchmarks and Analysis}
\label{sec:response_benchmarks}
In this section, we explore various benchmarks for evaluating the performance of language models in generating reliable responses. These benchmarks assess language understanding, factual accuracy, reliability, and the ability to provide timely information.

\textbf{(1) General Evaluation}. 
To comprehensively assess the language models' understanding capabilities across a wide range of scenarios, MMLU~\cite{MMLU} utilizes a multiple-choice format covering 57 different tasks, from basic mathematics to American history, computer science, and law. This benchmark spans evaluations in humanities, social science, and science, technology, engineering, and mathematics, providing a comprehensive and challenging test. It has been widely used in the evaluation of Large Language Models (LLMs) in recent years~\cite{llama, llama2, mistral_7b}.

Furthermore, BIG-bench~\cite{BIG-bench} introduces a large-scale and diverse benchmark designed to measure and understand the capabilities and limitations of LLMs across a broad range of tasks. Including 204 tasks contributed by 450 authors from 132 institutions, it covers areas such as linguistics, mathematics, and common sense reasoning. It focuses on tasks beyond the capabilities of language models, exploring how model performance and societal biases evolve with scale and complexity.

LLM-Eval~\cite{LLM-Eval} offers a unified multi-dimensional automatic evaluation method for open-domain dialogue of LLMs, eliminating the need for manual annotation. The performance of LLM-Eval across various datasets demonstrates its effectiveness, efficiency, and adaptability, improving over existing evaluation methods. The research also analyzes the impact of different LLMs and decoding strategies on the evaluation outcomes, underscoring the importance of selecting suitable LLMs and decoding strategies.

For Chinese, C-Eval~\cite{C-Eval} aims to comprehensively evaluate LLMs' advanced knowledge and reasoning capabilities in the Chinese context. It is based on a multiple-choice format, covering four difficulty levels and 52 different academic fields from secondary school to professional levels. C-Eval also introduces C-Eval Hard, a subset containing highly challenging subjects to test the models' advanced reasoning capabilities. Through evaluating state-of-the-art English and Chinese LLMs, C-Eval reveals areas where current models still fall short in handling complex tasks, guiding the development and optimization of Chinese LLMs.

\textbf{(2) Tool Evaluation}.
To assess the ability of language models to utilize tools, API-Bank~\cite{API-Bank} provides a comprehensive evaluation framework containing 73 APIs and 314 tool usage dialogs, along with a rich training dataset of 1,888 dialogs covering 1,000 domains to improve LLMs' tool usage capabilities. Experiments show that different LLMs perform variably in tool usage, highlighting their strengths and areas for improvement.

Later, ToolBench~\cite{ToolLLM} developed a comprehensive framework including a dataset and evaluation tools to facilitate and assess the ability of LLMs to use over 16,000 real-world APIs. It enhances reasoning capabilities by automatically generating diverse instruction and API usage scenario paths, introducing a decision tree based on depth-first search. ToolBench significantly enhances LLMs' performance in executing complex instructions and in their ability to generalize to unseen APIs. ToolLLaMA, an LLM fine-tuned from LLaMA~\cite{llama}, exhibits remarkable zero-shot capabilities and performance comparable to state-of-the-art LLMs like ChatGPT~\cite{chatgpt}.

\textbf{(3) Factuality Evaluation}.
TruthfulQA~\cite{TruthfulQA} measures the truthfulness of language models in answering questions. This benchmark consists of 817 questions covering 38 categories, including health, law, finance, and politics. This evaluation reveals that, even in optimal conditions, the truthfulness of model responses only reaches 58\%, in stark contrast to human performance at 94\%. Moreover, they proposed an automated evaluation metric named GPT-judge, which classifies the truthfulness of answers by fine-tuning the GPT-3~\cite{gpt3} model, achieving 90-96\% accuracy in predicting human evaluations.

HaluEval~\cite{halueval} is a benchmark for evaluating LLM illusions, constructed using a dataset containing 35K illusion samples, employing a combination of automated generation and manual annotation. This provides effective tools and methods for assessing and enhancing large language models' capabilities in identifying and reducing illusions. For Chinese scenarios, HalluQA~\cite{HalluQA} designs 450 meticulously selected adversarial questions to assess the illusion phenomenon in Chinese LLMs, covering multiple domains and reflecting Chinese culture and history, identifying two main types of illusions: imitative falsehoods and factual errors.

To evaluate the ability of LLMs to generate answers with cited text, ALCE~\cite{alce} builds an end-to-end system for retrieving relevant text passages and generating answers with citations. ALCE contains three datasets, covering different types of questions, and evaluates the generated text's quality from 'fluency', 'correctness', and 'citation quality' dimensions, combining human evaluation to verify the effectiveness of the evaluation metrics. The experimental results show that while LLMs excel at generating fluent text, there is significant room for improvement in ensuring content factual correctness and citation quality, especially on the ELI5 dataset where the best model was incomplete in citation support half of the time.

\textbf{(4) Real-Time Evaluation}.
RealTime QA~\cite{RealTimeQA} created a dynamic question-and-answer platform that regularly releases questions and evaluates systems weekly to ask and answer questions about the latest events or information. It challenges the static assumption of traditional QA datasets aiming for immediate application. Experiments based on LLMs like GPT-3 and T5 found that models could effectively update their generated results based on newly retrieved documents. However, when the retrieved documents failed to provide sufficient information, models tended to return outdated answers.

Furthermore, FreshQA~\cite{freshllm} evaluates large language models' performance in challenges involving time-sensitive and erroneous premise questions by creating a new benchmark containing questions of this nature. Evaluating various open and closed-source LLMs revealed significant limitations in handling questions involving rapidly changing knowledge and erroneous premises. Based on these findings, the study proposed a simple in-context learning method, FreshPrompt, significantly improving LLMs' performance on FreshQA by integrating relevant and up-to-date information sourced from search engines into the prompt.

\textbf{(5) Safety, Ethic, and Trustworthiness}.
To comprehensively evaluate the safety of LLMs, SafetyBench~\cite{SafetyBench} implements an efficient and accurate evaluation of LLMs' safety through 11,435 multiple-choice questions covering 7 safety categories in multiple languages (Chinese and English). The diversity of question types and the broad data sources ensure rigorous testing of LLMs in various safety-related scenarios. Comparing the performance of 25 popular LLMs, SafetyBench revealed GPT-4's significant advantage and pointed out the areas where current models need improvements in safety to promote the rapid development of safer LLMs.

For ethics, TrustGPT~\cite{TrustGPT} aims to assess LLMs' ethical performance from toxicity, bias, and value alignment, three key dimensions. The benchmark uses predefined prompt templates based on social norms to guide LLMs in generating content and employs multiple metrics to quantitatively assess the toxicity, bias, and value consistency of these contents. Experimental analysis revealed that even the most advanced LLMs still have significant issues and potential risks in these ethical considerations.

For trustworthiness, TrustLLM~\cite{TrustLLM} explores principles and benchmarks including truthfulness, safety, fairness, robustness, privacy, and machine ethics across six dimensions. Extensive experiments, including assessing 16 mainstream LLMs' performance on 30 datasets, found that trustworthiness usually positively correlates with functional effectiveness. While proprietary models typically outperform open-source models in trustworthiness, some open-source models like Llama2 showed comparable high performance.

These benchmarks provide important tools and metrics for evaluating and improving the capabilities of language models, contributing to the development of more accurate, reliable, safe, and timely GenIR systems. For further understanding of the evaluation works,~\cite{survey-factuality, survey-eval, survey_hallu_llm, dai2024unifying} offer more detailed introductions.

\section{Challenges and Prospects}
\label{sec:challenge}

This section discusses the key challenges faced in the fields of generative document retrieval and reliable response generation, as well as potential directions for future research.

\subsection{Challenges on Generative Document Retrieval}
\label{sec:challenge_gr}

\subsubsection{Scalability Issues}
\label{subsec:scalability_issues}
As extensively studied by~\cite{pradeep_how_does}, generative retrieval demonstrates significantly lower retrieval accuracy compared to dense retrieval when handling million-level document corpora in web search scenarios. Merely increasing the model size does not yield stable performance improvements. However, GR outperforms dense retrieval in document collections smaller than 300K, posing a question: What impedes GR methods from scaling to large document sizes? This issue encompasses several aspects:

\textbf{Training Data}. Current LLMs are pre-trained on huge datasets ranging from hundreds of billions to several trillion tokens, covering vast knowledge sources such as the internet, books, and news articles, consuming substantial computational power~\cite{llm-survey}. They are then extensively fine-tuned with high-quality, human-annotated data to achieve substantial generalization capabilities~\cite{t5, bart, InstructGPT, llama}. In contrast, generative retrieval (GR) models often begin with a pre-trained language model and are fine-tuned on labeled data comprising <query, DocID> pairs, which does not sufficiently prepare them to fully grasp GR tasks. For numeric-based DocIDs, the models, having not encountered these numbers in their pre-training phase, tend to rote memorize the DocIDs seen during training, struggling to predict unseen ones effectively. Similarly, if text-based DocIDs fail to precisely represent the documents, the model also tend to rote learning.

A potential solution is to create a large-scale pre-training dataset for generative retrieval on a general corpus, possibly including a variety of common DocIDs such as URLs, titles, and numerical sequences. We can utilize instructions to distinguish generation targets for various DocIDs. Then we can pre-train a GR model from scratch, the model can understand generative retrieval across diverse domains. This method could bridge the gap between language model pre-training data and GR tasks, enhancing the generalization ability of GR models across different corpora.

\textbf{Training Method}.
As described in Section~\ref{sec:gr_training}, existing training methods explore various training objectives, including seq2seq training, learning DocID, and ranking capabilities. Other methods involve knowledge distillation~\cite{understand-dsi}, reinforcement learning~\cite{genrrl}, etc. Is there a better training method to enable GR models to master generating DocID ranking lists? For example, RLHF~\cite{rlhf} has been effectively used to train LLMs~\cite{InstructGPT, llama2}, though at a high cost. Exploring RLHF in the GR field is also worthwhile.

\textbf{Model Structure}.
As discussed in Section~\ref{sec:gr_structure}, most current GR models are based on encoder-decoder Transformers structures~\cite{dsi, nci, ultron}, such as T5~\cite{t5} and BART~\cite{bart}. Some GR methods like CorpusLM~\cite{corpuslm} have experimented with a decoder-only structure of the LLM Llama2~\cite{llama2}, requiring more training computational power but not significantly improving performance. Research is needed to determine which structure is more suitable for generative retrieval. Additionally, whether increasing model and data size could lead to emergent phenomena similar to those observed in LLMs~\cite{Wei2022Emergent, Schaeffer2023AreEmergent} is also a promising research direction.

\subsubsection{Handling Dynamic Corpora}
\label{subsec:handling_dynamic_corpora}
Real-world applications often involve dynamically changing corpora, such as the web and news archives, where incremental learning is essential. However, for language models, indexing new documents inevitably leads to forgetting old ones, posing a challenge for GR systems. Existing methods like DSI++~\cite{dsi++}, IncDSI~\cite{incdsi}, CLEVER~\cite{clever}, and CorpusBrain++~\cite{corpusbrain++} propose solutions such as experience replay, constrained optimization, incremental product quantization, and continual generative pre-training frameworks to address incremental learning issues. Yet, these methods have their specific applicable scenarios, and more effective and universally applicable incremental learning strategies remain a key area for exploration.

\subsubsection{Document Identifier}
\label{subsec:document_representation}
Accurately representing a document with high-quality DocIDs is crucial for generative retrieval.

For example, the KILT dataset based on the Wikipedia corpus, which includes 5.9 million documents, demonstrates optimistic retrieval performance for GR methods using titles as DocIDs~\cite{genre, corpusbrain, corpuslm}. This is because each document in Wikipedia has a unique manually annotated title that represents the core entity discussed in that page. However, in the web search scenario, such as in the MS MARCO dataset~\cite{ms_marco}, many documents lack a unique title, are overlapping, and the titles do not accurately represent the core content of the documents. Thus, GR performance significantly declines in the MS MARCO corpus of 8.8 million passages.

Therefore, how to construct high-quality titles (or other types of DocIDs) in general corpora, similar to those in Wikipedia, that not only accurately represent documents but also are lightweight, is a critical factor for implementing GR methods and warrants in-depth research.

\textbf{Text or Numeric?}
As discussed in Section~\ref{sec:docid_design}, current methods include text-based and numeric-based DocIDs, each with their advantages and disadvantages. Text-based DocIDs effectively leverage the linguistic capabilities of pre-trained generative language models and offer better interpretability. Numeric-based DocIDs can utilize dense retriever embeddings to obtain semantic DocID sequences; they can also complement dense retrievers to achieve synergistic benefits.

However, to ensure good generalization ability of GR models without extensive pre-training, it is essential to utilize the inherent pre-trained parameters of the model. Coherent textual DocIDs can naturally leverage this aspect, but they also need to capture key document semantics and maintain linguistic sequence characteristics. Numeric DocIDs, however, do not offer this advantage. Thus, as mentioned in \ref{subsec:scalability_issues}, extensive pre-training is necessary to enable models to fully understand the meanings behind these numerical strings, which is a costly endeavor.

\textbf{Do We Need a Unique ID for Each Document?}
Most current GR methods use a unique DocID to uniquely identify a document. However, as the number of documents in a corpus increases, maintaining a unique DocID becomes increasingly challenging. Even if a unique DocID is maintained, it is difficult to differentiate significantly from other DocIDs semantically, leading to reduced retrieval precision. Some methods, such as using sub-string as DocIDs~\cite{seal, ugr}, have proven effective. These methods utilize the FM-Index~\cite{fm_index} to ensure the generated sub-string exists in the corpus and use the number of generated sub-strings in different documents to rank documents, demonstrating good performance and generalization ability.

However, since this method is based on FM-Index, its inference latency is high, which is an issue that needs addressing. Furthermore, exploring other more efficient alternatives to FM-Index and even considering not using constrained search but freely generating a DocID sequence followed by a lightweight matching and scoring module to efficiently return a document ranking list are also worthy of exploration.

\subsubsection{Efficiency Concerns}
\label{subsec:efficiency_concerns}
Current GR methods generally rely on constrained beam search to generate multiple DocID sequences during inference, resulting in high latency. This is particularly severe when returning 100 or more documents, with latencies reaching several hundred milliseconds~\cite{nci}, which is unacceptable for low-latency IR systems. Therefore, designing more efficient inference methods is crucial. To reduce inference latency, the length of the DocID sequence should not be too long; 16 tokens or fewer is an efficient range. This necessitates designing DocIDs that are precise and concise enough to represent documents while maintaining performance and improving efficiency. Additionally, developing more efficient decoding strategies is a valuable research direction for the future.

\subsubsection{Multi-modal Generative Retrieval}
\label{subsec:challenge_multi_modal}

Existing multi-modal generative retrieval models aim to retrieve images by converting each image in the collection into a unique sequence that serves as its identifier. A language model is then employed to predict these image identifiers, enabling effective image retrieval. However, there are still potential areas for future optimization: 
\textbf{(1) Image Representation:} Developing advanced image representation techniques is essential for enhancing the performance of multi-modal generative retrieval. These techniques should capture the key features of an image within its identifier sequence.
\textbf{(2) End-to-end Training:} Existing methods perform image representation and image identifier prediction separately for generative retrieval. Exploring how to train these two tasks in a fully end-to-end manner is also worth investigating.
\textbf{(3) Extend to Additional Modalities:} Current multi-modal generative retrieval methods predominantly focus on text and image modalities. Expanding these approaches to incorporate additional modalities such as audio and video presents a valuable research opportunity.

\subsection{Challenges on Reliable Response Generation}
\label{sec:challenge_response}

\subsubsection{Improving Accuracy and Factuality}
\label{subsec:improving_accuracy}
In GenIR systems, ensuring content accuracy and factuality is crucial. To achieve this, as mentioned in Section~\ref{sec:response_generation}, there are two main areas of improvement:

\textbf{Internal Knowledge Memorization.}
Firstly, training stronger generative models is critical for building reliable GenIR systems. Various commercial LLMs continue to progress, utilizing vast training data and computational resources, but exploring better model structures is also worthwhile. Recent research such as Retentive Networks~\cite{RetentiveNetwork}, Mamba~\cite{mamba}, and others, have shown potential to challenge the performance and efficiency of Transformers~\cite{transformer}. However, whether these can scale and truly surpass Transformer-based LLMs in generation quality is still an open question. Moreover, what types of training data and methods can consistently produce models capable of generating high-quality, reliable text also deserve thorough investigation and summary. The mechanisms by which language models recall knowledge during inference are not yet clear and need to be fully understood to better serve user information needs.

\textbf{External Knowledge Enhancement.}
As described in Section~\ref{sec:retrieval_augmentation}, retrieval-augmented generation is an effective method widely applied in LLMs. However, there is still room for improvement. (1) For example, whether inserting retrieved documents directly into generative models via prompts is the best method, or if there are better ways, such as inputting embeddings~\cite{cepe}, needs exploration. (2) Additionally, whether models can autonomously decide whether to perform retrieval~\cite{selfdc, SlimPLM}, and when in the generation process to perform it~\cite{flare}. (3) Third, in dialogue scenarios, enhancing RAG models to better utilize long conversational history is also worth further exploration~\cite{2410_conv_search_survey}.

Tool-augmented generation, as discussed in Section~\ref{sec:tool_augmentation}, is also a popular method for endowing LLMs with fine-grained world knowledge and performing complex tasks. Recent research has raised questions, such as "Should tools always be used?"~\cite{What_Are_Tools_Anyway}. More specifically, whether the performance improvements brought by using tools justify the extra computational costs incurred during model training or the inference costs during testing. Existing work mainly focuses on task accuracy, but studying the cost-effectiveness of these methods is also a valuable topic.

\subsubsection{Real-time Properties of GenIR Systems}
\label{subsec:real_time}
Timeliness is critical for GenIR systems, as well as traditional IR systems, to provide users with the most up-to-date information. However, since the knowledge of pre-trained generative models is fixed after training, methods like retrieval and tool augmentation are needed to acquire new external knowledge. Research on real-time knowledge acquisition remains limited, making it a valuable area for investigation.

Moreover, continually relying on outdated knowledge from language models is inadequate, as models cannot comprehend the significance of given contexts or backgrounds in the current era, thus reducing the reliability of the generated content. Therefore, updating the information in language models while avoiding the forgetting of existing knowledge, such as through continual learning~\cite{study_continual_leaning, survey_continual_learning}, knowledge editing~\cite{survey_knowledge_edit_LLMs, Finetune_or_Retrieval, emnlp2023editing_LLMs, survey_knowledge_edit_NN}, etc., is a topic worth further exploring.

\subsubsection{Bias and Fairness}
\label{subsec:bias}
Since LLMs are often trained on large, unfiltered datasets, GenIR systems may propagate stereotypes and biases present in the data regarding race, culture, and other aspects~\cite{survey_Bias_and_Fairness}. Researchers have explored various methods to enhance the fairness of generated content during training data selection, training methods, generation techniques, and rewriting phases. However, biases have not been eradicated and require a thorough understanding of the mechanisms by which generative models produce biases, to design methods to solve them and build fair GenIR systems that further the practical application of GenIR.

\subsubsection{Privacy and Security}
\label{subsec:data_privacy}
Firstly, the content generated by GenIR systems risks plagiarism~\cite{Dien2023AI_Plagiarism, kenthapadi2023responsibleai}. Studies such as~\cite{Extracting_Training_Data_from_LLMs, PLM_Personal_Information} indicate that pre-trained language models can reproduce large segments of their training data, leading to inadvertent plagiarism and causing academic dishonesty or copyright issues. On one hand, legal regulations regarding the copyright of AI-generated content will gradually emerge and evolve. On the other hand, technical research aimed at reducing plagiarism by generative models, such as generating text with correct citations~\cite{metzler2021rethinking, EvaluatingVerifiability, Huang2023Citation_A_Key}, is a promising research direction for reliable GenIR that has received increasing attention in recent years.

Moreover, due to the unclear mechanisms of memory and generation in pre-trained language models, GenIR systems inevitably return unsafe content. For example,~\cite{TheSecretSharer, Extracting_Training_Data_from_LLMs, 2409_trust_rag_survey} show that when attacked, LLMs may return private information of users seen in training data. Therefore, understanding the mechanisms by which LLMs recall training data and designing effective defense mechanisms to enhance security are crucial for the widespread use of GenIR systems. Additionally, developing effective detection methods for content generated by LLMs is essential for enhancing the security of GenIR systems~\cite{survey_llm_detection}.

\subsection{Unified Framework}
\label{sec:unified_framework}
This article discusses two mainstream forms of GenIR: generative document retrieval and reliable response generation. However, each approach has its advantages and limitations. Generative document retrieval still returns a list of documents, whereas the reliable response generation model itself cannot effectively capture document-level relationships. Therefore, integrating these two approaches is a promising research direction.

\subsubsection{Unified Framework for Retrieval and Generation}
\label{subsec:unified_retrieval_generation}
Given that both generative retrieval and downstream generation tasks can be based on generative language models, could a single model perform both retrieval and generation tasks? Indeed, it could.

Current attempts, such as UniGen~\cite{unigen}, use a shared encoder and two decoders for GR and QA tasks respectively, and show superior performance on small-scale retrieval and QA datasets. However, they struggle to generalize across multiple downstream tasks and to integrate with powerful LLMs. Additionally, CorpusLM~\cite{corpuslm} uses a multi-task training approach to obtain a universal model for GR, QA, and RAG. Yet, merely merging training data does not significantly improve retrieval and generation performance, and CorpusLM remains limited to the Wikipedia corpus. Facing a broader internet corpus presents significant challenges.

In the future, can we construct a large search model (LSM) that allows an LLM to have the capability to generate DocIDs and reliable responses autonomously? Even LSM could decide when to generate DocIDs to access the required knowledge before continuing generation. Unlike the large search model defined in~\cite{wang2023LargeSearchModel}, which unifies models beyond the first-stage retrieval (such as re-ranking, snippet, and answer models), we aim to integrate the first-stage retrieval as well, enabling the LSM to fully understand the meaning of retrieval and its connection with various downstream generation tasks.

\subsubsection{Towards End-to-End Framework for Various IR Tasks}
\label{subsec:end2end_ir_framework}
Metzler et al.~\cite{metzler2021rethinking} envisioned an expert-level corpus model that not only possesses linguistic capabilities but also understands document-level DocIDs and knows the sources of its own knowledge. Such a model could not only solve the issue of hallucinations common in traditional language models but could also generate texts with references pointing to the source documents, thus achieving a reliable end-to-end GenIR model. By understanding DocIDs and knowledge sources, this end-to-end system could also perform additional IR tasks, such as returning the main content of a document given its DocID or returning other related document DocIDs, as well as enabling multi-lingual and multi-modal retrieval.

Current methods, as discussed in this GenIR survey, primarily focus on generative document retrieval (GR) and response generation as separate entities. GR models excel at comprehending document identifiers at the document-level, while downstream models demonstrate powerful task generation capabilities. However, existing methods face challenges when it comes to effectively integrating these two generative abilities, limiting the overall performance and effectiveness of the GenIR system. The integration of these generative abilities in a seamless and efficient manner remains a key challenge in the field. 

In the future, we can design training methods that align knowledge and DocIDs and construct high-quality training datasets for generating answers with references, to train such an end-to-end GenIR model. Achieving this goal remains challenging and requires the collaborative efforts of researchers to contribute to building the next generation of GenIR systems.

\section{Conclusion}
\label{sec:conclusion}

In this survey, we explore the latest research developments, evaluations, current challenges, and future directions in generative information retrieval (GenIR). We discuss two main directions in the GenIR field: generative document retrieval (GR) and reliable response generation. Specifically, we systematically review the progress of GR covering model training, document identifier design, incremental learning, adaptability to downstream tasks, multi-modal GR, and generative recommendation systems; as well as advancements in reliable response generation in terms of internal knowledge memorization, external knowledge enhancement, generating responses with citations, and personal information assistance. 
Additionally, we have sorted out the existing evaluation methods and benchmarks for GR and response generation. We organize the current limitations and future directions of GR systems, addressing scalability, handling dynamic corpora, document representation, and efficiency challenges. Furthermore, we identify challenges in reliable response generation, such as accuracy, real-time capabilities, bias and fairness, privacy, and security. We propose potential solutions and future research directions to tackle these challenges.
Finally, we also envision a unified framework, including unified retrieval and generation tasks, and even building an end-to-end framework capable of handling various information retrieval tasks. Through this review, we hope to provide a comprehensive reference for researchers in the GenIR field to further promote the development of this area.

\appendix

\section{Details for Evaluation}
\label{app:metrics}

\subsection{Evaluation Metrics for Generative Document Retrieval}
\label{app:gr_metrics}

\textbf{Recall}. Recall is a metric that measures the proportion of relevant documents retrieved by the search system. For a given cutoff point $k$, the recall $\text{Recall}@k$ is defined as:
\begin{equation}
    \text{Recall}@k = \frac{1}{|Q|} \sum_{q=1}^{|Q|} \frac{ret_{q,k}}{rel_q},
\end{equation}
where $|Q|$ is the number of queries in the set, $ret_{q,k}$ is the number of relevant documents retrieved for the $q$-th query within the top $k$ results, and $rel_q$ is the total number of relevant documents for the $q$-th query.

\textbf{R-Precision}. R-Precision measures the precision at the rank position $R$, which corresponds to the number of relevant documents for a given query $q$. It is calculated as:
\begin{equation}
    \text{R-Precision} = \frac{ret_{q,R}}{rel_q},
\end{equation}
where $ret_{q,R}$ is the number of relevant documents retrieved within the top $R$ positions, and $R$ is equivalent to $rel_q$.

\textbf{Mean Reciprocal Rank (MRR)}. MRR reflects the average rank position of the first relevant document returned in the search results. It is computed as follows:
\begin{equation}
    \text{MRR} = \frac{1}{|Q|} \sum_{q=1}^{|Q|} \frac{1}{\text{rank}_q},
\end{equation}
where $\text{rank}_q$ is the rank of the first relevant document returned for the $q$-th query.

\textbf{Mean Average Precision (MAP)}. MAP calculates the average precision across multiple queries. It considers the exact position of all relevant documents and is calculated using the following formula:
\begin{equation}
    \text{MAP} = \frac{1}{|Q|} \sum_{q=1}^{|Q|} \left( \frac{1}{rel_q} \sum_{k=1}^{n_q} \text{P}@k \times I(q,k) \right),
\end{equation}
where $\text{P}@k$ is the precision at cutoff $k$, and $I(q,k)$ is an indicator function that is 1 if the document at position $k$ is relevant to the $q$-th query and 0 otherwise.

\textbf{Normalized Discounted Cumulative Gain (nDCG)}. nDCG takes into account not only the relevance of the documents returned but also their positions in the result list. It is defined by:
\begin{equation}
    \text{DCG}@k = \sum_{i=1}^{k} \frac{2^{\text{rel}_i} - 1}{\log_2(i+1)},
\end{equation}
\begin{equation}
    \text{nDCG}@k = \frac{\text{DCG}@k}{\text{IDCG}@k},
\end{equation}
where $\text{rel}_i$ represents the graded relevance of the $i$-th document, $\text{DCG}@k$ is the discounted cumulative gain, and $\text{IDCG}@k$ represents the maximum possible $\text{DCG}@k$.

\subsection{Benchmarks for Generative Document Retrieval}
\label{app:gr_benchmarks}

\textbf{MS MARCO}. MS MARCO (Microsoft Machine Reading Comprehension) is a large-scale dataset developed by Microsoft for evaluating machine reading comprehension, retrieval, and question-answering capabilities within web search contexts. It comprises two primary benchmarks:

\begin{itemize}
    \item \textbf{Document Ranking}: This benchmark includes approximately 3.2 million documents derived from real user queries extracted from Microsoft Bing's search logs. Each query is paired with annotated relevant documents, facilitating the evaluation of retrieval accuracy and scalability.
    \item \textbf{Passage Ranking}: Containing around 8.8 million passages, this benchmark focuses on more granular retrieval tasks, assessing the system's ability to identify relevant information at the passage level.
\end{itemize}

The diversity of question types and document genres in MS MARCO aims to mimic complex web search scenarios, making it a pivotal resource for testing the robustness and effectiveness of GR systems.

\textbf{Natural Questions (NQ)}. Natural Questions (NQ) is a question-answering dataset introduced by Google, utilizing Wikipedia as its foundational corpus. It encompasses approximately 3.2 million documents, each corresponding to a Wikipedia page. The dataset includes a wide array of natural user queries along with their respective answers extracted directly from web pages in Google search results. NQ is designed to evaluate the retrieval performance of GR systems in addressing real-world, information-seeking questions, emphasizing the ability to understand and retrieve precise answers from a vast knowledge base.

% \textbf{TriviaQA}. TriviaQA is a comprehensive QA dataset that includes a substantial number of factual questions and answers sourced from diverse origins such as Wikipedia and web-based QA forums. The dataset comprises approximately 660,000 evidence documents, providing a rich and varied set of data for evaluating the generalization capabilities of information retrieval systems. TriviaQA is particularly useful for assessing how well GR methods can handle different domains and a wide range of question types, thereby testing the adaptability and robustness of retrieval algorithms.

\textbf{KILT}. KILT (Knowledge Intensive Language Tasks) is an extensive benchmark dataset that integrates five categories of knowledge-intensive tasks, including:

\begin{itemize}
    \item \textbf{Fact Checking}: Utilizing datasets like FEVER, KILT assesses the system's ability to verify factual claims against a knowledge base.
    \item \textbf{Entity Linking}: Incorporates datasets such as AIDA CoNLL-YAGO, WNED-WIKI, and WNED-CWEB to evaluate the linking of entities mentioned in text to their corresponding entries in a knowledge base.
    \item \textbf{Slot Filling}: Includes T-REx and Zero Shot RE datasets to test the system's ability to populate predefined slots with relevant information extracted from the text.
    \item \textbf{Open-Domain QA}: Combines datasets like Natural Questions, HotpotQA, TriviaQA, and ELI5 to evaluate the retrieval and comprehension capabilities of the system in answering open-ended questions.
    \item \textbf{Dialogue}: Utilizes the Wizard of Wikipedia dataset to assess the system's performance in maintaining informative and coherent dialogues based on retrieved knowledge.
\end{itemize}

KILT employs Wikipedia as its primary corpus, consisting of approximately 5.9 million wiki pages. The benchmark aims to evaluate the effectiveness of information retrieval systems in handling complex language tasks that require extensive background knowledge and the ability to integrate information across multiple domains.

\textbf{TREC Deep Learning Track 2019 \& 2020}. The TREC Deep Learning Tracks for 2019 and 2020 are specialized evaluation campaigns focusing on the application of deep learning techniques to enhance the efficiency and effectiveness of information retrieval systems. The primary tasks in these tracks include:

\begin{itemize}
    \item \textbf{Document Ranking}: Assessing the ability of retrieval systems to rank entire documents based on their relevance to a given query.
    \item \textbf{Passage Ranking}: Evaluating the system's capability to identify and rank specific passages within documents that are most relevant to the query.
\end{itemize}

\bibliographystyle{ACM-Reference-Format}
\bibliography{main}

\end{document}